\documentclass[10pt]{aastex}
\usepackage{emulateapj5,apjfonts}
\usepackage{graphicx}
\slugcomment{{\sc Accepted to ApJ:} September 20, 2004}
\newcommand{\begit}{\begin{itemize}}
\newcommand{\enit}{\end{itemize}}
\newcommand{\begen}{\begin{enumerate}}
\newcommand{\enen}{\end{enumerate}}

\newcommand       \be           {\begin{equation}}
\newcommand       \ee           {\end{equation}}
\newcommand       \bea          {\begin{eqnarray}}
\newcommand       \eea          {\end{eqnarray}}
\newcommand      \fgas          {f_{g_{\,0.1}}}
\newcommand       \kms		{\,{\rm km \,\, s}^{-1}}

\newcommand       \ergs		{\,{\rm erg \,\, s}^{-1}}
\newcommand{\beqa}{\begin{eqnarray}} 
\newcommand{\eeqa}{\end{eqnarray}}

\def\mpy{\rm \ M_\odot \ {\rm yr^{-1}}}

\begin{document}

\title{On the Maximum Luminosity of Galaxies \& Their Central Black
Holes: Feedback From Momentum-Driven Winds}

\author{Norman Murray\altaffilmark{1,2,3}, Eliot Quataert\altaffilmark{4}, \& Todd A.~Thompson\altaffilmark{5,6}}

\altaffiltext{1}{Canada Research Chair in Astrophysics}
\altaffiltext{2}{Visiting Miller Professor, The University of California, Berkeley}
\altaffiltext{3}{Canadian Institute for Theoretical Astrophysics, 60 St.~George Street, University of Toronto, Toronto,
ON M5S 3H8, Canada; murray@cita.utoronto.ca}
\altaffiltext{4}{
Astronomy Department 
\& Theoretical Astrophysics Center, 601 Campbell Hall, 
The University of California, Berkeley, CA 94720; 
eliot@astro.berkeley.edu}
\altaffiltext{5}{Hubble Fellow}
\altaffiltext{6}{Astronomy Department 
\& Theoretical Astrophysics Center, 601 Campbell Hall, 
The University of California, Berkeley, CA 94720; 
thomp@astro.berkeley.edu}

\begin{abstract}

We investigate large-scale galactic winds driven by momentum
deposition. Momentum injection is provided by (1) radiation pressure
produced by the continuum absorption and scattering of photons on
dust grains and (2) supernovae (momentum injection by supernovae is
important even if the supernovae energy is radiated away).  
Radiation can be produced by a starburst or AGN activity. 

We argue that momentum-driven winds are an efficient mechanism for
feedback during the formation of galaxies.  We show that above a
limiting luminosity, momentum deposition from star formation can expel
a significant fraction of the gas in a galaxy.  The limiting,
Eddington-like luminosity is $L_{\rm M}\simeq(4f_g c/G)\,\sigma^4$,
where $\sigma$ is the galaxy velocity dispersion and $f_g$ is the gas
fraction; the subscript M refers to momentum driving.  A starburst
that attains $L_{\rm M}$ moderates its star formation rate and its
luminosity does not increase significantly further.  We argue that
ellipticals attain this limit during their growth at $z \gtrsim 1$ and
that this is the origin of the Faber-Jackson relation.  We show that
Lyman break galaxies and ultra-luminous infrared galaxies have
luminosities near $L_{\rm M}$.  Since these starbursting galaxies account
for a significant fraction of the star formation at $z\gtrsim1$, this
supports our hypothesis that much of the observed stellar mass in
early type galaxies was formed during Eddington-limited star
formation.

Star formation is unlikely to efficiently remove gas from very small
scales in galactic nuclei, i.e., scales much smaller than that of a
nuclear starburst.  This gas is available to fuel a central black hole
(BH).  We argue that a BH clears gas out of its galactic nucleus when
the luminosity of the BH itself reaches $\approx L_{\rm M}$.  This
shuts off the fuel supply to the BH and may also terminate star
formation in the surrounding galaxy.  As a result, the BH mass is
fixed to be $M_{\rm BH}\simeq (f_g \kappa_{\rm es}/\pi G^2)\sigma^4$,
where $\kappa_{\rm es}$ is the electron scattering opacity. This limit
is in accord with the observed $M_{\rm BH}-\sigma$ relation.

\end{abstract}


\section{Introduction}

Large elliptical galaxies in the local universe exhibit a relation
between their luminosity $L$ and the depth of their gravitational
potential wells (as measured by their stellar velocity dispersion
$\sigma$) of the form $L\propto \sigma^4$, a result first noted nearly
thirty years ago (Faber \& Jackson 1976). More recently it was found
that most nearby early-type galaxies (ellipticals and spiral bulges)
contain massive black holes, and that the mass $M_{\rm BH}$ of the
hole scales as $M_{\rm BH}\propto\sigma^4$ (Ferrarese \& Merritt 2000;
Gebhardt et al.~2000; Tremaine et al.~2002).  If these black holes
radiate near their Eddington limit, their luminosity would also
satisfy $L\propto\sigma^4$. It would be remarkable if this
correspondence with the Faber-Jackson (FJ) relation is a
coincidence.

In an apparently unrelated phenomenon, nearby starburst galaxies,
which are generally spirals, but also include dwarf irregulars and
dwarf ellipticals, are seen to drive large-scale galactic outflows
(Heckman, Armus, \& Miley 1990; Martin 1999; Heckman 2000; Strickland
2004; Martin 2004). More distant starburst galaxies include Lyman Break Galaxies
(LBGs; Steidel et al.~1996) and SCUBA sources (e.g., Smail, Ivison, \&
Blain 1997).  These also show evidence for large-scale outflows (e.g.,
Pettini et al. 2000; Adelberger et al. 2003).  The SCUBA sources have
infrared luminosities as large as $10^{13}L_\odot$, making them
Ultra-Luminous Infrared Galaxies (ULIRGs; e.g., Genzel \& Cesarsky
2000).  The space density and mass of the ULIRGs suggest that they are
the progenitors of present day massive ellipticals.

In this article we argue that all of these phenomena are intimately
related; they result directly from a limit on the luminosity of
massive self-gravitating gas-rich objects set by momentum deposition
in the interstellar medium.  We show that significant momentum
injection into the ISM of star-forming galaxies may be accomplished by
two sources: radiation pressure from the continuum absorption and
scattering of photons on dust grains, and
supernovae.\footnote{Haehnelt (1995) also considered some properties
of feedback by momentum deposition during galaxy formation, focusing
on radiation pressure from Lyman edge photons, rather than dust or
supernovae.}  The photons may come either from the starburst itself
or from a central massive black hole. Supernovae have often been
considered as an energy source for thermal pressure-driven galactic
winds. Less consideration has been given to supernovae as a source of
momentum flux into the ISM; unlike energy, the momentum
supernovae deposit cannot be radiated away.  

Starburst galaxies both locally and at high redshift are typically
highly reddened (e.g., Heckman, Armus, \& Miley~1990; Meurer et
al.~1995; Sanders \& Mirabel 1996; Adelberger \& Steidel 2000;
Calzetti 2001; Genzel et al.~2004). Optical depths to UV/IR photons
may easily exceed unity, suggesting that a large fraction of the
momentum created by star formation is available to drive an
outflow. As noted above, a central AGN provides an alternative source
of photons. We show that either source can drive a galactic wind.
Previous authors have considered the possibility that dust itself is
expelled from galaxies by radiation pressure, particularly in the
context of enriching the IGM with metals (Davies et al. 1998; Aguirre
1999; Aguirre et al. 2001abc).  We argue that, as in models of
dust-driven stellar winds (e.g., Netzer \& Elitzur 1993), the dust and
gas are hydrodynamically coupled and thus that the dust can drag the
gas out of the galaxy.

This paper is organized as follows. We begin by considering the
general properties of momentum-driven galactic winds in
\S\ref{section:momentum}. We show via an Eddington-like argument that
there exists a limiting starburst luminosity above which a large
fraction of the gas in a galaxy can be expelled.  When the gas is
optically thick, the limiting luminosity is given by
\be 
\label{eq:basic result}
L_{\rm M}=\frac{4f_gc}{G}\,\sigma^4,
\ee
where $f_g$ is the fraction of mass in gas.\footnote{A potentially related empirical limit on
the surface brightness of starburst galaxies has been described by Meurer et al.~(1997).}
Scoville (2003) has considered an analogous optically thin Eddington limit in setting
the maximum luminosity per unit mass in star-forming galaxies.
In \S\ref{section:energy}, we contrast the properties of momentum-driven
winds with those of energy-driven winds that are more typically
invoked in the galactic context (Chevalier \& Clegg 1985; Heckman,
Armus, \& Miley 1990).  We also show that the dynamics of cold gas
entrained in a hot thermal wind is analogous to that of the
momentum-driven outflows considered in \S2 (we elaborate on this point
in an Appendix).

With equation (\ref{eq:basic result}) in hand, we focus on the
importance of this limit for setting the observed properties of
elliptical galaxies.  In \S\ref{section:starburst} we present
observational evidence that the star formation rates required to reach
our limiting luminosity are realized during the formation of massive
galaxies at high redshift. These high star formation rates are
probably initiated by galaxy mergers. It follows that star formation
in elliptical galaxies self-regulates via momentum deposition. We show
that this model can account for the Faber-Jackson relation between the
current luminosity and velocity dispersion of early-type galaxies.

We also summarize data showing that the most luminous galaxies at any
$\sigma$ and redshift, not just massive ellipticals, roughly satisfy
equation (\ref{eq:basic result}). We argue that this implies that energy
deposition by supernovae is not efficient at globally halting star
formation, even in small galaxies.  This is in contrast to the
conventional picture in the galaxy formation literature (Dekel \& Silk
1986).

We then consider the relative role of AGN and star formation in
driving large scale galactic winds (\S\ref{section:agn}).  We provide
observational evidence that the most luminous AGN have luminosities
$\approx L_{\rm M}$.  This supports a model in which accretion onto
AGN self-regulates in a manner similar to that of star formation on
galactic scales.  When the AGN luminosity (and BH mass) exceeds a
critical value, the AGN clears gas out of the galactic nucleus,
shutting off its own fuel supply.  This can account for the observed
$M_{\rm BH}-\sigma$ relation.  Our treatment of self-regulated black
hole growth is similar to that of King (2003; see also Silk \& Rees
1998; Haehnelt, Natarajan, \& Rees 1998; Blandford 1999; Fabian 1999;
Fabian et al.~2002).

Finally, in \S\ref{section:conclude} we summarize our results and
discuss further implications of momentum-driven galactic winds.  

\section{Momentum-Driven Galactic Winds}

In this section we review the basics of momentum-driven
winds.  In \S\ref{section:energy} we contrast the scalings
derived here for momentum deposition with the corresponding
relations for energy-driven galactic outflows.

\subsection{Preliminaries}
\label{section:preliminary}

We take as a model for the gravitational
potential that of an isothermal sphere with gas density and mass
profiles given by 
\be 
\rho(r) = {f_g \sigma^2 \over 2 \pi G r^2}
\label{rhor}
\ee 
and 
\be 
M_g(r)={2 f_g\sigma^2 r\over G},
\label{mor}
\ee 
where $\sigma$ is the velocity dispersion and $f_g$ is the gas
fraction.  We assume that $f_g$ is a constant throughout this work.

The goal of this paper is both to elucidate the physics of
momentum-driven galactic winds, and to discuss the applicability of
such outflows to rapidly star-forming galaxies at high redshift.  For
the latter purpose, it is convenient to consider several physical
scales characterizing galaxies.  We follow the treatment of Mo, Mao,
\& White (1998).  The virial radius of the dark matter halo is given
by 
\beqa
R_{\rm V}&=&\sqrt{2}\sigma\Big/[10H(z)] \nonumber \\
&\sim&285\,\,{\rm kpc}\,\,\sigma_{200}\,h^{-1}[H_0/H(z)],
\eeqa 
where $H(z)$ is the Hubble constant at redshift $z$ where the halo is
formed, $H_0=100\,h\kms{\rm Mpc}^{-1}$, and
$\sigma_{200}=\sigma/200\kms$.  The dynamical timescale on the scale
$R_{\rm V}$ is
\be 
\tau_{\rm Dyn}^{\rm V}=R_{\rm V}\big/\sigma\sim1.4\,\,{\rm Gyr}\,\,h^{-1}[H_0/H(z)].
\label{taudynv}
\ee 
The total gas mass within a dark matter halo of dispersion
$\sigma$ is $\approx M_g(R_V)$.  Using equation (\ref{mor}) this yields
\be 
M_g = 5 \times 10^{11}\, M_\odot\,\fgas\,\sigma_{200}^3\,h^{-1}[H_0/H(z)],
\label{mgtot}
\ee
where $\fgas=f_g/0.1$.

Although the mass in dark matter is distributed out to the virial
radius $R_V$, the baryons cool and condense in the dark matter halo
and are thus significantly concentrated with respect to the dark
matter. The ``disk'' radius characterizing the baryons is
related to the virial radius by \beqa R_{\rm D}&=&\frac{1}{\sqrt{2}}
\lambda R_{\rm V} \nonumber \\ &\sim&10\,\,{\rm
kpc}\,\,\lambda_{0.05}\,\sigma_{200}\,h^{-1}[H_0/H(z)],
\label{rd}\eeqa where $\lambda_{0.05}=\lambda/0.05$ is the spin
parameter of the dark matter halo (Bullock et al.~2001).  The
dynamical timescale on the scale $R_{\rm D}$ is then 
\be \tau_{\rm
Dyn}^{\rm D}=R_{\rm D}/\sigma\sim50\,\,{\rm
Myr}\,\,\lambda_{0.05}\,h^{-1}[H_0/H(z)].
\label{taudynd}
\ee 
Note that both $\tau_{\rm Dyn}^{\rm V}$ and $\tau_{\rm Dyn}^{\rm
D}$ are independent of $\sigma$ and are fixed fractions of the Hubble
time, $H(z)^{-1}$.

There is a final length and time scale that is not easily deduced from
parameters of the dark matter, namely that characterizing a starburst.
Observations of systems ranging from local dwarf starbursts to ULIRGs
at high redshift show that star formation can be distributed on scales
ranging from $\sim 100$ pc to several kpc.  It is important to note
that although the dynamical time of the starburst region might be
rather short (e.g., $\approx 5$ Myrs for a nuclear burst on kpc
scales), the duration of the star formation activity ($\equiv
\tau_{\rm SB}$) can be significantly longer.  In particular, for
starbursts triggered by a major merger, the duration of the burst may
be set by the duration of the merger, which is several dynamical times
$\tau_{\rm Dyn}^{\rm D}$ (eq.~[\ref{taudynd}]).

\subsection{Momentum Injection \& The Mass Loss Rate}
\label{section:momentum}

The maximal mass loss rate $\dot{M}_{\rm W}$ of a momentum-driven outflow
from an object with total momentum deposition rate $\dot{P}$ is given by
\be 
\dot{M}_{\rm W}\,V_\infty \approx
\dot{P},
\label{pmom}
\ee 
where $V_\infty$ is the terminal velocity of the wind. We show below
that for galaxies $V_\infty \sim \sigma$.  We consider two primary sources of
momentum deposition in driving large-scale galactic outflows:
supernovae and radiation pressure from the central starburst or AGN.
In the former case, assuming that each SN produces $\approx 10$ M$_\odot$
of material moving at $v \approx 3000\kms$, we estimate a net momentum
deposition rate of
\be 
\dot{P}_{\rm SN}\sim2\times10^{33}\,\,
{\rm g\,cm\,s^{-2}}\,
\left({\dot{M}_\star \over 1 \,\ {\rm M_\odot \ yr^{-1}}}\right),
\label{pdotsn}
\ee 
where $\dot{M}_\star $ is the star formation rate and we assume 1
supernova per 100 years per $M_\odot$ yr$^{-1}$ of star
formation.\footnote{Shell-shell collisions in a spatially homogeneous distribution of 
supernova remnants will cancel momentum at shell-shell interfaces. Hence,
only outside the region where the distribution can be described
as homogeneous will there be a significant outward radial momentum
flux.  Therefore, the estimate of equation \ref{pdotsn} should be regarded
as an upper limit to the net momentum injection 
available for driving a wind.}
This momentum deposition by supernovae occurs even if the
kinetic energy of the explosion is efficiently radiated away.  Note
also that winds from massive stars can provide a momentum flux
comparable to that provided by supernovae (Leitherer et al. 1999).

In the case of radiation pressure from a nuclear starburst or AGN, in
the point-source, single-scattering limit, $\dot{P}=L/c$, where $L$ is
the luminosity of the central radiating object, and $L/c$ is the total
momentum flux.  Although both a starburst and an AGN may contribute to
the galaxy luminosity, we consider only the starburst contribution to
the total luminosity ($L_{\rm SB}$) in this section.  We explore the
role of AGN in \S\ref{section:agn}.

We can compare $\dot{P}_{\rm SN}$ with $L_{\rm SB}/c$ for the
starburst by writing $L_{\rm SB}=\epsilon\dot{M}_\star c^2$.
Examination of the starburst models of Leitherer et al.~(1999) and
Bruzual \& Charlot (2003) indicates that for a Salpeter
IMF,\footnote{The efficiency $\epsilon$ depends on the low-mass
cutoff of the IMF, $m_l$, as $\epsilon \propto m_l^{0.35}$.}
$\epsilon\sim10^{-3}\equiv\epsilon_3$, or that
\be 
L_{\rm SB}/c\sim2\times10^{33}\,\,{\rm
g\,\,cm\,\,s^{-2}} \, \epsilon_3 \, 
\left({\dot{M}_\star \over {\rm 1 \ M_\odot \ yr^{-1}}}\right).
\label{pdotsb}
\ee 
Comparing eqs.~(\ref{pdotsn}) and (\ref{pdotsb}) we see that the net
momentum deposited by supernova explosions is roughly the same as that
deposited by stars.  With this in mind, we write equation (\ref{pmom})
as
%
\be 
\dot{M}_{\rm W} V_\infty\approx L_{\rm SB}/c.
\label{mom} 
\ee 
If the driving mechanism is pure radiation pressure from a central
starburst or AGN this equality is only appropriate if the flow has
optical depth ($\tau$) of order unity.  More generally, given an
optical depth $\tau$, ``$\tau L/c$" replaces ``$L/c$".

Since both $\dot{P}_{\rm SN}$ and $L_{\rm SB}$ are proportional to the
star formation rate $\dot{M}_\star $, equation (\ref{mom}) immediately
implies that in a momentum-driven galactic wind the mass loss rate is
proportional to the star formation rate:
\be 
\dot{M}_{\rm
W}\sim\dot{M}_\star \,\left(\frac{\epsilon c}{V_\infty}\right) =
\dot{M}_\star \,\left(\frac{300 \ \epsilon_3 \ {\rm km \
s^{-1}}}{V_\infty} \right). \label{wind} 
\ee 
This implies that, for $V_\infty \sim \sigma \sim 200\kms$, $\dot M_{\rm W}
\sim \dot M_\star$.


\subsection{Wind Dynamics: Optically Thick Limit}

We approximate the gas surrounding a point source with
luminosity $L$ as a spherical optically-thick shell.  Ignoring gas
pressure, the momentum equation for the gas can be written as
\be 
{dP\over dt}=M_g(r){dV\over dt}=-{GM(r)M_g(r)\over r^2}+{L(t)\over c},
\label{thickmom}
\ee 
Using equation (\ref{mor}) we see that if $L(t)$ is less than the critical luminosity
\be 
L_{\rm M}=\frac{4 f_g c}{G}\,\,\sigma^4,
\label{lmom}
\ee 
where the subscript ``M" on the limiting luminosity $L_{\rm M}$ stands
for momentum-driven, then the effective gravity is reduced by the
momentum deposition of the radiation, but the motion of the gas is
inward toward the central point source. For $L\gtrsim L_{\rm M}$, the
gas moves outward in a radiation pressure-driven outflow.  Taking
$L(t)$ constant in time in equation (\ref{thickmom}) implies
\be 
\frac{dV}{dt}=\frac{GM(r)}{r^2}\left[\frac{L}{L_{\rm
M}}-1\right]=\frac{2\sigma^2}{r}\left[\frac{L}{L_{\rm M}}-1\right]. \label{thickmom2}
\ee 
Taking equation (\ref{thickmom2}) as the momentum equation for a
time-independent optically thick wind (not a shell) and integrating,
we obtain
\be 
V(r)=2\sigma\sqrt{\left[\frac{L}{L_{\rm
M}}-1\right]\ln\left(\frac{r}{R_0}\right)},
\label{vthick}
\ee 
where $R_0$ is the initial radius of the outflow and we have
neglected $V(R_0)$.  For $L$ of a few times $L_{\rm M}$ and distances
as large as several $R_{\rm V}(\gg R_0)$, the asymptotic velocity does
not exceed several times $\sigma$, i.e., $V_\infty \approx3 \sigma$.

When $L \gtrsim L_{\rm M}$, the momentum injected by star formation is
sufficient to blow out {\it all} of the gas in the galaxy. Taking
$\fgas=f_g/0.1$ and $\sigma_{200}=\sigma/200$ km s$^{-1}$ yields
\be 
L_{\rm M}\simeq3\times10^{46}\,\,\,{\rm
erg\,\,s^{-1}}\,\,\,\fgas\sigma_{200}^4.
\label{lmoms}
\ee 
The star formation rate corresponding to $L_{\rm M}$ is then 
\be 
\dot{M}_\star =L_{\rm M}/\epsilon c^2\simeq500\,\,{\rm M_\odot
\,\,yr^{-1}}\,\,\fgas\epsilon_3^{-1}\sigma_{200}^4.  \label{mdotstar}
\ee 
From equation (\ref{mom}), the outflow rate is
\be 
\dot{M}_{\rm W}\simeq500\,\,{\rm
M_\odot\,\,yr^{-1}}\,\,\,\fgas\sigma_{200}^3.
\label{mmom}
\ee 
Equations~(\ref{lmom}) and (\ref{mmom}) imply that $\dot{M}_{\rm
W}\propto \sigma^3 \propto L^{3/4}$.

It is worth considering whether the luminosity $L_{\rm M}$ can
plausibly be obtained in a starburst.  In \S4 we provide observational
evidence that it is, while here we present a simple theoretical
estimate. The maximum star formation rate in a dark matter potential
well can be estimated by first assuming that the gas builds up on a
scale $\sim R_D$ without much star formation, \`{a} la Mo, Mao \& White
(1998).  Mergers and interactions between galaxies can then
efficiently convert this gas into stars on a timescale $\sim \tau^{\rm
D}_{\rm Dyn} \sim R_{\rm D}/\sigma$, the merger timescale.  This can
in principle generate a star formation rate approaching
\beqa 
\dot{M}_\star^{\rm max}=\frac{M_g}{\tau_{\rm Dyn}^{\rm D}}&=& \frac{2^{3/2}f_g}{\lambda
G}\,\sigma^3 \nonumber \\ &\sim&10^4\,\,{\rm
M_\odot\,\,yr^{-1}}\fgas\lambda_{0.05}^{-1}\sigma_{200}^3, 
\eeqa 
producing a maximum luminosity of 
\beqa 
L_{\rm max}&=&\epsilon\dot{M}_\star^{\rm max}c^2 \nonumber \\
&\sim& 6\times10^{47}\,\,{\rm
erg\,\,s^{-1}}\,\,\epsilon_{3}\fgas\lambda_{0.05}^{-1}\,\sigma_{200}^3.
\label{lmax}
\eeqa 
With this estimate for the maximum star formation rate, there is a
critical $\sigma \equiv \sigma_{\rm max}$ above which a starburst
cannot generate the luminosity required to expel the gas (i.e.,
$L_{\rm max} \lesssim L_{\rm M}$):
\be 
\sigma_{\rm max}=\frac{\epsilon c}
{\sqrt{2}\lambda}\sim4000\,\,{\rm km\,\,s^{-1}}\,\,\epsilon_3\,
\lambda^{-1}_{0.05}.
\label{sigmacritd}
\ee 
This estimate suggests that star formation rates required to produce a
luminosity $\sim L_{\rm M}$ can plausibly be achieved, particularly in
mergers when stored gas is converted into stars on a timescale $\sim
\tau^{\rm D}_{\rm Dyn}$.  The actual value of $\sigma_{\rm max}$ is,
of course, quite uncertain because it depends on the efficiency of
star formation and ambiguities in defining the dynamical timescale for
the starburst.  Equation (\ref{sigmacritd}) may well be an
overestimate because the gas mass available for star formation at any
time may be significantly less than the total mass $M_g$ in 
equation (\ref{mgtot}).


\subsection{Wind Dynamics: Optically Thin Limit}

So far we have assumed that the flow is optically thick at the base of
the wind.  This is certainly appropriate in the case of pure momentum
driving by supernovae ($\dot{P}_{\rm SN}$).  However, if $\dot{P}$ is
provided by the luminosity of the starburst or AGN, then we must
distinguish between the optically thick and optically
thin limits.  If the spherical shell surrounding a point source with
luminosity $L$ is optically thin, again ignoring gas pressure, the
momentum equation for the shell is 
\be 
{dV\over dt}=-{GM(r)\over r^2}
+{\kappa L\over 4\pi r^2 c},
\label{thinmom}
\ee 
where the optical depth $\tau=\kappa M_g(r)/4\pi r^2$ and $\kappa$ is
the opacity (per unit mass of gas).  Thus, the condition on the
luminosity for the shell to move outward is the classical (optically
thin) Eddington result, 
\beqa 
L_{\rm SB}\gtrsim L_{\rm Edd}&=&{4\pi
GM(r)c\over\kappa}={8\pi c \over\kappa}\,\sigma^2 r \nonumber \\
&\approx&10^{46}\,\, \ergs\sigma_{200}^2\,r_{\rm kpc}\,\kappa_{100}^{-1},
\label{lthin}
\eeqa 
where $\kappa_{100}=\kappa/100$ cm$^{2}$ g$^{-1}$ and $r_{\rm kpc}=r/1$ kpc.
The velocity profile for a steady-state optically thin wind can be obtained by integrating equation (\ref{thinmom}),
\be
V(r)=2\sigma\sqrt{R_{\rm g}\left({1\over R_0}-{1\over r}\right) + \ln(R_0/r)},
\label{vthin}
\ee where \be R_{\rm g}\equiv {\kappa L\over 8\pi c \sigma^2}={L\over
L_{\rm Edd}(R_0)}\, R_0.  \label{rg} \ee The gas, close to $R_0$,
initially accelerates. When it reaches $R_{\rm g}$ it begins to
decelerate, eventually stopping if the galaxy is much larger than the
initial launch radius $R_0$. In order for the gas to reach ten times
its initial radius, we require $L/L_{\rm Edd}\sim3$.

Equations (\ref{thinmom})-(\ref{rg}) also apply to optically thick
{\it clouds} that fill only a fraction of the volume in the galaxy (in
contrast to the optically thick {\it shell} considered in \S2.3).  For
a cloud of mass $M_c$ and area $A_c$, the force per unit mass a
distance $r$ from the luminosity source is $A_cL/(4 \pi r^2 c M_c)$.
This is identical to the force in equation (\ref{thinmom}) with $\kappa
\rightarrow A_c/M_c$.  In this case the Eddington limit can be
rewritten as \be L_{\rm Edd} \approx 2 \times 10^{45}\,\,
\ergs\sigma_{200}^2\,r_{\rm kpc}\,N_{21} \label{leddcloud} \ee where
$N_{21} = N_H/10^{21} {\rm cm^2}$ is the hydrogen column and we have
rewritten the mass per unit area of the cloud as $M_c/A_c = (4/3) m_p
N_H$. More properties of the acceleration of optically thick clouds
are considered in \S\ref{section:entrainment} and the Appendix.

\subsection{The Critical Opacity \& Dust Production}

The difference between the limiting luminosity derived in the
optically thick case ($L_{\rm M}$; eq.~\ref{lmom}) and that derived in
the optically thin case ($L_{\rm Edd}$; eq.~\ref{lthin}) is important.
The dividing line between these physical regimes is given by a
critical opacity $\kappa_{\rm crit}$ above which the gas is optically
thick.  To estimate $\kappa_{\rm crit}$, we assume that all of the gas
in a galaxy is concentrated on the scale $R_{\rm D}$ (see eq.~[\ref{rd}]).
The condition $\tau \gtrsim 1$ then requires $\kappa \gtrsim
\kappa_{\rm crit}$ with 
\beqa 
\kappa_{\rm crit} &\approx& \frac{\pi G
\lambda^2}{\sqrt{2}\sigma f_g 10 H(z)} \nonumber \\ &\simeq& 6 \ {\rm
cm^2\,\,g^{-1}}
\,\,\lambda_{0.05}^2\,\,\sigma_{200}^{-1}\,\,\fgas^{\,-1}\,\,h^{-1}\frac{H_0}{H(z)}.
\label{kc}
\eeqa Note that this estimate applies both if the gas is distributed
spherically on a scale $\sim R_{\rm D}$ or if it is in a disk ($\tau$
is then the vertical optical depth through the disk).  If $\kappa
<\kappa_{\rm crit}$ (particularly likely at lower redshift where $f_g
\ll 0.1$ or in a small galaxy with $\sigma_{200} \ll 1$), then the
optically thin limit obtains (eq.~\ref{lthin}).  Conversely, if
$\kappa>\kappa_{\rm crit}$, then the optically thick limit obtains and
the limiting luminosity required to drive the gas mass to infinity via
momentum deposition is given by equation (\ref{lmom}).

The critical opacity obtained in equation (\ref{kc}) is much larger than
the electron scattering opacity ($\kappa_{es}\simeq0.38$ cm$^2$
g$^{-1}$), but it is easily provided by continuum dust absorption and
scattering of photons (e.g.~Draine \& Lee 1984).  Dust opacity can
be in the range of several hundred cm$^2$ g$^{-1}$ and is responsible
for the high reddening observed in both local (e.g.~Calzetti 2001;
Heckman, Armus, \& Miley~1990; Meurer et al.~1995; Lehnert \& Heckman
1996) and distant starbursting galaxies including ULIRGs (Sanders \&
Mirabel 1996) and LBGs (e.g.~Adelberger \& Steidel 2000).  Note that
$\kappa > \kappa_{\rm Crit}$ can be obtained even for very subsolar
metallicities ($\lesssim 0.1$ solar) suggesting that the momentum
driven outflows considered here may be important even for the
formation of relatively `primordial' galaxies.

It is worth considering how and in what quantity dust is created in
young galaxies.  Dust can be created in AGB stars, but the timescale
to do so is $\sim 1\,{\rm Gyr}$, long compared to the
duration of a starburst, and so may not dominate the production of
dust in young galaxies. Instead, Kozasa et al.~(1989), Todini \&
Ferrara (2001), and Nozawa et al.~(2003) show that supernovae can
produce $M_{\rm Dust}\sim0.5$ M$_\odot$ of dust per supernova,
depending on the progenitor metallicity and mass.  To order of
magnitude, with ${\cal L}$ the scale of the system, the number of
supernovae required to yield $\tau_{\rm Dust}\sim1$ in the volume
$(4\pi/3){\cal L}^3$ is 
\be 
N_{\rm SN}\sim
10^6\left(\frac{\cal{L}}{2\,\,{\rm kpc}}\right)^{2}
\left(\frac{0.5\,\,{\rm M_\odot}}{M_{\rm Dust}}\right)
\left(\frac{10^4\,\,{\rm cm^2\,\,g^{-1}}}{\kappa_{\rm Dust}}\right),
\label{nsn} 
\ee 
where $\kappa_{\rm Dust}$ is the opacity in units of cm$^2$ per gram
of {\it dust} (note that throughout the rest of this paper $\kappa$ is
expressed in units of cm$^2$ per gram of {\it gas}; we use
$\kappa_{\rm Dust}$ in equation \ref{nsn} because the result can then
be expressed independent of the gas mass or gas-to-dust ratio).  For a
supernova rate of $10^{-2}$ yr$^{-1}$ for every solar mass per year of
star formation, the timescale for supernovae to generate $\tau_{\rm
Dust}\sim1$ is $\sim10^8$ yr $\dot{M}_\star^{-1}$.  This timescale is
reasonably short and so we expect that the required opacity may be
produced either by quiescent star formation or during a starburst
itself.  For example, in a large starburst with $\dot{M}_\star\sim100$
M$_\odot$ yr$^{-1}$, $\tau_{\rm Dust}\sim1$ is reached in just
$\sim10^6$ yr (or soon after the first supernovae occur).


\subsection{The Coupling Between Dust \& Gas}

The mean free path for scattering of dust and gas is simply
$\lambda=(n \sigma_{\rm dg})^{-1}$, where $n$ is the gas number
density and $\sigma_{\rm dg}$ is the dust-gas scattering cross
section.  Since we require that order unity of the dust momentum be
imparted to the gas, the quantity of interest is $\lambda_{\rm
M}=\lambda(m_{\rm D}/m_p)$, where $m_{\rm D}$ is the mass of an
individual dust grain.  For a grain of radius $a$ and a geometric
cross section, we estimate \be \lambda_{\rm M}\simeq10\,\,{\rm
pc}\,\,a_{0.1}\, \rho_{3} \, n^{-1}_{1},
\label{lambdam}
\ee where $a_{0.1}=a/(0.1 \mu{\rm m})$, $n_1$ is the number density of
gas, normalized to one particle per cm$^3$, and $\rho_{3}$ is the mass
density of an individual dust grain, normalized to three gram per
cm$^3$.  To assess the hydrodynamical coupling of dust and gas we
compare $\lambda_{\rm M}$ with the radial scale $R$ in the galaxy.  We
leave to future work a detailed multi-fluid model of dust-driven
galactic winds (as in models of winds from cool stars).

In a sufficiently low density phase of the ISM, it is possible to have
$\lambda_{\rm M} > R$ and thus for dust and gas to be hydrodynamically
decoupled.  In this case dust could be expelled from a galaxy without
the gas (e.g., Davies et al. 1998).  It seems, however, more likely
that the dust is present in the cold, dense phase of the ISM with $n_1
\gtrsim 1$, in which case $\lambda_M \ll R$ and the dust efficiently
shares the radiative momentum it receives with the gas.  As the flow
moves outwards, however, the density of gas decreases and
$\lambda_{\rm M}$ may eventually exceed the radius $R$.  To estimate
the radial scale at which this happens ($R_{\rm dg}$), we use
$n=\dot{M}_{\rm W}/(4\pi m_p R^2 V)$ to estimate the gas density (with
$\dot M_{\rm W}$ from eq.~\ref{mmom} and $V \sim \sigma$).  Combining
with equation (\ref{lambdam}) we find that $\lambda_{\rm M} > R$ at
\be R_{\rm dg}=\frac{3}{4\pi}\frac{\sigma^2 f_g}{\rho_{\rm D}a
G}\sim150\,\,{\rm kpc}\,\, \rho^{-1}_3\, a^{-1}_{0.1}\,
\sigma_{200}^{2}\, \fgas.  \ee Because $R_{\rm dg}$ is significantly
larger than the scale on which the outflow is accelerated, we expect
the dust and gas to be well coupled in the acceleration region and
thus that the gas can be blown out of the galaxy with the dust.  This
estimate of $R_{\rm dg}$ is probably conservative because the dust may
primarily be in cold gas clouds whose density will not decrease as
$R^{-2}$ as for the continuous wind considered above.  Note also that
near the source of UV photons the dust grains will be charged.  This
will further increase the coupling of the dust to the gas, both
because of Coulomb collisions (Draine \& Salpeter 1979) and because
the Larmor radius of the dust will be sufficiently small that the dust
is magnetically coupled to the gas.


\section{Energy-Driven Galactic Winds}
\label{section:energy}

Several properties of winds generated by energy deposition are
different from those driven by momentum deposition.  To highlight the
differences, we briefly review the physics of the former in the
galactic context (see, e.g., Chevalier \& Clegg 1985 for analytic
solutions).  

Given a total energy deposition rate $\dot{E}$ (erg s$^{-1}$), one may
estimate the mass loss rate of an energy deposition-driven wind by
equating the asymptotic kinetic energy loss rate with $\dot{E}$
\be 
\frac{1}{2}\dot{M}_{\rm W}V_\infty^2 \approx \dot{E}.
\label{en}
\ee 
For a starburst galaxy the energy injection is provided by
supernovae and winds from massive stars, with comparable energy from
each source (Leitherer et al.~1999).  We focus on supernovae here.
Assuming each supernova yields an energy of $\sim10^{51}$ erg, the
total energy deposition rate from supernovae is 
\beqa 
\dot{E}_{\rm
SN} &=& \xi E_{\rm SN}\Gamma_{\rm SN} \sim \xi E_{\rm SN}\dot{M}_\star
f_{\rm SN} \nonumber \\ &\sim&3\times10^{40}\,\,{\rm erg
\,\,s^{-1}}\,\,\xi_{0.1}\,\left({\dot{M}_\star \over {\rm 1 \,M_\odot
\,yr^{-1}}}\right) ,
\label{edotsn}
\eeqa 
where $\xi$ is the efficiency of energy transfer to the ISM
($\xi_{0.1}=\xi/0.1$ implies 10\% efficiency), $\Gamma_{\rm SN}$ is
the number of supernovae per unit time, and $f_{\rm SN} \approx
10^{-2}$ is the number of supernovae per solar mass of star formation.
The efficiency $\xi$ with which SN energy is transferred to the ISM is
uncertain and depends on, e.g., the density of the ISM. Thornton et
al.~(1998) showed that supernova remnants typically radiate at least
90\% of their energy during their evolution.  Hence, only $\sim$10\%
may be efficiently thermalized in the ISM.  We normalize 
$\xi$ to this value, but emphasize that it is uncertain.  

Comparing equation (\ref{edotsn}) with equation (\ref{pdotsb}) we see that,
absent radiative losses ($\xi = 1$), $\dot{E}_{\rm SN}$ can be written
simply in terms of the starburst luminosity as $\dot{E}_{\rm SN} \sim
10^{-2} L_{\rm SB}$.  The factor of $100$ appearing in this
relationship comes from the fact that a typical massive star releases
$\sim10^{53}$ erg in luminous energy over its lifetime, whereas it
deposits $\sim 10^{51}$ erg during its supernova (e.g.~Abbott 1982).

Equations (\ref{en}) and (\ref{edotsn}) can be combined to
give an expression for the mass loss rate in energy-driven winds;
\beqa
\dot M_{\rm W} & \approx & \dot M_\star \left(\xi \epsilon 10^{-2} c^2
\over V_\infty^2 \right) \nonumber \\
&\approx & \dot M_\star \xi_{0.1} \epsilon_3 \
\left({ 300 \ \kms \over
V_\infty}\right)^2. 
\label{mdoten} 
\eeqa
For fiducial numbers this estimate is similar to our estimate of the
mass loss rate in momentum driven winds (eq.~\ref{wind}).  However, 
eqs.~(\ref{mdoten}) and (\ref{wind}) differ in two important ways.
First, if the supernova energy is efficiently radiated away ($\xi \ll 1$),
equation (\ref{mdoten}) predicts a mass loss rate much less than equation (\ref{wind}).  
Second, momentum-driven winds yield the scaling $\dot{M}_{\rm W} \propto 
\dot{M}_\star/V_\infty$, whereas energy-driven winds predict $\dot
M_{\rm W} \propto \dot{M}_\star/V^2_\infty$.  This difference in scaling 
may be observationally testable.

We estimate the energy injection required to unbind the gas in a
galaxy by requiring that $\dot{E}_{\rm SN} \tau_{\rm Dyn} \gtrsim
E_{\rm Bind}$, where $E_{\rm Bind}=GMM_g/r$ and $\tau_{\rm Dyn}\sim
r/\sigma$.  This yields $\dot{E}_{\rm SN}\gtrsim 4f_g\sigma^5/G$.
Rewriting this in terms of the corresponding starburst luminosity
gives 
\beqa 
L_{\rm E}&\sim&\frac{100}{\xi}\frac{4f_g}{G}\,\sigma^5 \nonumber \\
&\sim&2\times10^{46}\,\,\,{\rm
erg\,\,s^{-1}}\,\,\,\fgas\,\xi_{0.1}^{-1}\,\sigma_{200}^5,
\label{edot}
\eeqa 
where the subscript ``E'' stands for ``energy-driven,'' in contrast to
$L_{\rm M}$ (eq.~\ref{lmom}).  Equation (\ref{edot}) is a criterion to
``blow away'' all of the gas of the galaxy (following the nomenclature
of De Young \& Heckman 1994).  In a disk galaxy, supernovae may ``blow
out'' in the direction perpendicular to the disk (even for $L_{\rm
SB}$ less than $L_{\rm E}$), but in this case very little of the gas
mass of the galaxy will be affected (e.g., De Young \& Heckman 1994).
In fact, the numerical simulations of MacLow \& Ferrara (1999) and
Strickland \& Stevens (2000) find that supernova energy thermalized in
the ISM can be efficiently vented when the remnants break out of the
galactic disk.  However, very little of the mass in the galaxy is
actually blown away.


\subsection{Entrainment}
\label{section:entrainment}

A hot wind can in principle sweep up and entrain embedded clouds of
cold gas, driving them out of the galaxy by ram pressure.  The cold
gas may also be shock heated and evaporated by the hot flow. The
energy-driven limit considered in the previous subsection is
appropriate when most of the cold gas is shock heated and incorporated
into the hot flow (and radiative losses are small).  In the opposite
limit, in which the cold gas retains its identity, the dynamics of the
cold gas is analogous to that of a momentum-driven wind because it is
pushed out by the ram pressure of the hot gas.  In the Appendix we
show this explicitly by demonstrating that the ram pressure and
radiation pressure forces on cold clouds are typically comparable
(eq. [\ref{ramvsrad}]).  We also derive both the optically thin
(eq. [\ref{leddcloud}]) and optically thick (eq. [\ref{lmom}])
Eddington luminosities discussed in \S\ref{section:momentum} using ram
pressure as the acceleration mechanism (rather than radiation
pressure); see, in particular, equations (\ref{mdothot}) and
(\ref{lmoment}).

Distinguishing which of these two mechanisms actually dominates the
acceleration of cold gas is an important but difficult problem.  It
is, in particular, unclear whether embedded clouds can actually
survive entrainment in a hot flow.  Calculations show that the clouds
are typically destroyed in a few cloud crossing times (e.g., Klein,
McKee, \& Colella 1994; Poludnenko, Frank, \& Blackman 2002), though
considerable cloud material can be accelerated to high velocity in the
process.  In the Appendix we focus on the acceleration of cloud
material to highlight the analogy between ram pressure and radiation
pressure driving of cold gas, but the acceleration, ablation, and
destruction of the clouds likely go hand in hand.

\subsection{Comparing Momentum-Driven \& \\ Energy-Driven Winds}
\label{section:compare}

Equation (\ref{edot}) defines the starburst luminosity above which
energy injection by supernovae is sufficient to unbind all of the gas
in the galaxy.  Comparing this limiting luminosity with the
corresponding expression for momentum-driven winds ($L_{\rm M}$;
eq.~[\ref{lmom}]), we see that $L_{\rm M}\sim L_{\rm E}$ at a critical
velocity dispersion \be \sigma_{\rm crit}=\frac{\xi}{100} \,c
\sim300\,\,{\rm km\,\,s^{-1}}\xi_{0.1}.
\label{sigmacrit}
\ee For $\sigma<\sigma_{\rm crit}$, $L_{\rm E}<L_{\rm M}$ and one
might expect energy deposition via supernovae to dominate feedback on
the ISM.  By contrast, for $\sigma>\sigma_{\rm crit}$, $L_{\rm
M}<L_{\rm E}$ and momentum deposition dominates and is sufficient to
blow all of the gas out of the galaxy (this is true regardless of
whether the primary source of momentum deposition is radiation pressure or
supernovae).  Note that because of the many simplifications made in
deriving $L_{\rm M}$ and $L_{\rm E}$, the actual value for
$\sigma_{\rm crit}$ is only accurate to order of magnitude; it is also
very sensitive to assumptions about the efficiency with which
supernova energy is thermalized in the ISM.  In
\S\ref{section:starburst} we show that dwarf starbursting galaxies
violate equation (\ref{edot}) by several orders of magnitude.  This
suggests that in practice $\sigma_{\rm crit}$ is quite small,
significantly smaller than the nominal value in equation
(\ref{sigmacrit}).

The scale $\sigma_{\rm crit}$ sets a lower bound on the range of
$\sigma$ over which the luminosity limit $L_{\rm M}$ is applicable.
There is also an upper bound.  In eqs.~(\ref{lmax}) and
(\ref{sigmacritd}) we estimated (very crudely) the maximum star
formation rate and luminosity attainable in a starburst and the
$\sigma_{\rm max}$ above which a system cannot generate a luminosity
$\sim L_{\rm M}$.

Figure \ref{plot:lss} illustrates these bounds and the limiting
luminosities schematically.  The two limiting luminosities ($L_{\rm
E}\propto\sigma^5$ and $L_{\rm M}\propto\sigma^4$) as well as the
maximum attainable luminosity ($L_{\rm max}\propto\sigma^3$
eq.~[\ref{lmax}]) are sketched as a function of $\sigma$.  The limit
$L_{\rm M}$ is applicable in the region $\sigma_{\rm
crit}<\sigma<\sigma_{\rm max}$.  For reference, the Faber-Jackson
relation for elliptical galaxies and bulges is also sketched.  It has
a lower ``zero-point'' than $L_{\rm M}$, but the same dependence on
$\sigma$, $L_{\rm FJ}\propto\sigma^4$. We discuss this correlation and
its relation to $L_{\rm M}$ in detail in the next section but note
here that deviations from the FJ relation are possible for
$\sigma<\sigma_{\rm crit}$ and $\sigma>\sigma_{\rm max}$.


\section{Starburst Galaxies \& The Faber-Jackson Relation}
\label{section:starburst}

In this section we apply the idea of ``Eddington limited'' star
formation to star forming galaxies at high redshift.  The basic
scenario is as follows.  The luminosity of a nuclear starburst
increases as it forms stars.  When the luminosity increases to $L_{\rm
M}$ (eq.~[\ref{lmom}]) the starburst drives gas out of the galactic
potential and
regulates its luminosity to $\approx L_{\rm M}$ (a similar idea has been proposed 
by Elmegreen 1983 and Scoville et al.~2001 for what determines the mass
and luminosity of individual star clusters; Scoville 2003 discussed a limit
directly analogous to our optically thin limit $L_{\rm Edd}$, eq.~[\ref{lthin}]).
We argue
that this self-regulation determines the total number of stars formed
in a given dark matter potential well.  First, we describe observations
showing that star formation rates sufficient to produce $L\approx
L_{\rm M}$ do occur in star forming galaxies at both high and low
redshifts. Then we show that if all early type galaxies went through
such a star formation episode at $z \gtrsim 1$, self-regulation at
$\approx L_{\rm M}$ can explain the Faber-Jackson relation.

\subsection{The Maximum Luminosity $L_{\rm M}(\sigma)$: Observations}

Figure \ref{plot:ls} shows the
luminosity as a function of the velocity dispersion for a sample of
high star formation rate galaxies drawn from the literature.  We also
plot the expression for $L_{\rm M}$ (eq.~[\ref{lmom}]) for three values
of the gas fraction $f_g=1$, 0.1, and 0.01.  These different curves
should be taken to include both plausible variations in the gas
fraction (which changes in time), as well as uncertainty in the value
of $L_{\rm M}$.  The latter arises because the total momentum
deposition rate may be somewhat larger than just $L/c$ since
contributions from supernovae, stellar winds, and starburst photons
are all comparable. In addition, photons may be absorbed several times
as they exit the starburst region and the galaxy.

In collecting the data in Figure \ref{plot:ls}, we attempted to find
representative examples of the highest star formation rate galaxies at
a variety of $\sigma$ (see Tables \ref{tab:ls} and \ref{tab:ls3} for
details).  This includes dwarf galaxies (Mateo 1998; Martin 1998),
LBGs at $z \approx 2$ (Erb et al. 2003) and $z \approx 3$ (Pettini et
al. 2001), ULIRGs locally (Genzel et al. 2001) and at high redshift
(Neri et al. 2003; Genzel et al.~2003; Tecza et al.~2004), galaxies
from the CFRS survey at $z\approx0.6$ (Lilly et al. 1996; M{\'
a}llen-Ornelas et al.~1999), and a sample of local starbursts (Heckman
et al.~2000).  For the local starbursts we chose a sample of systems
that clearly show evidence for outflowing {\it cold} gas.  
Even though many systems  fall significantly below the $L_{\rm M}$ curve,
radiation pressure is sufficient to generate an outflow of cold gas
comparable to what is observed because 
the radiation pressure force on individual gas clouds can exceed
gravity even if $L \ll L_{\rm M}$ (since the latter criteria refers to
blowing out {\it all} of the gas in the galaxy).  We will discuss this
in more detail in a future paper.  

The data in Figure \ref{plot:ls} are necessarily heterogeneous, and there are
uncertainties in both luminosities and velocity dispersions, but this
compilation illustrates several important points.  First, the simple
momentum driving limit given by equation (\ref{lmom}) does provide a
reasonable upper limit to the luminosity of observed starbursting
systems.  The fact that some systems fall below this limit is, of
course, no surprise.  They might simply not have star formation rates
sufficient to reach $L_{\rm M}$, or they might be observed somewhat
after the peak star formation episode (which is, after all, where
systems spend most of their time; see Fig.~\ref{plot:decay1}).

It is also worth stressing that the upper envelope to the observed
luminosity as a function of $\sigma$ is incompatible with the simple
limit based on energy feedback from supernovae, which predicts $L_{\rm
E} \propto \sigma^5$ (\S\ref{section:energy}; eq.~[\ref{edot}]).  In
particular, the low $\sigma$ systems in Figure \ref{plot:ls}  have luminosities well
in excess of the energy limit given in equation (\ref{edot}).  This implies
that either the efficiency of transferring supernova energy to the ISM
is very low (e.g., $\xi \sim 10^{-2}$) or else supernovae do not
globally halt star formation by ejecting most of the gas (e.g.,
because the supernovae ``blow out'' of the galactic plane; de Young \&
Heckman 1994).  In either interpretation, this argues for $\sigma_{\rm
crit} \lesssim 20$ km s$^{-1}$ (see Fig.~\ref{plot:lss}), in which
case momentum injection may dominate the global mass loss in many
starbursting systems.

For the purposes of this paper, perhaps the most interesting feature
of Figure 2 is that starbursting galaxies at high redshift have
luminosities reasonably close to $L_{\rm M}$.  This includes both
LBGs, ULIRGs, and galaxies drawn from the CFRS redshift survey of
Lilly et al. (1996).  We suggest that this is not a coincidence, but
is instead evidence that star formation at high redshifts self-regulates;
when the starburst reaches a luminosity $\sim
L_{\rm M}$, the galaxy drives a powerful wind that limits the
available gas supply and thus the star formation rate.  This feedback
mechanism regulates the luminosity of the starburst and ultimately
helps set the stellar mass of the galaxy.

The $z \gtrsim 1$ galaxies shown in Figure 2 are representative of
systems that have been used to study the star formation history of the
universe (e.g., Madau et al. 1996; Steidel et al. 1999).  It is known
that integrating the inferred star formation history over redshift can
account reasonably well for the total stellar mass density observed at
$z = 0$ (Madau et al. 1998).
The fact that
many of the individual systems that comprise the `Madau' plot
have $L \sim L_{\rm M}$ thus suggests that a significant fraction of the
stellar mass in the universe has been built up through starbursts that
self-regulate by momentum-driven galactic winds.  We show in the next
section that if this hypothesis is correct, it can account for the
Faber-Jackson relation.

A direct test of our hypothesis is that rapidly star forming galaxies
at high redshift should drive powerful galactic winds.  Powerful winds
are seen in LBGs (e.g., Pettini et al. 2000; Adelberger et
al. 2003).  It is, however, difficult to isolate the physical
mechanism responsible for driving such outflows. One prediction of the
momentum-driven wind model is that observed outflows should have a
momentum flux $\dot M_{\rm W} V_\infty$ comparable to that of the starburst,
$L/c$.  This can be rewritten as (eq.~[\ref{wind}]) $\dot M_{\rm W} \approx
\dot M_\star (c \epsilon/V_\infty)$. This prediction is difficult to test
because it is hard to reliably measure the mass outflow rate $\dot
M_{\rm W}$.
The best case so far at high redshift is
probably the gravitationally lensed LBG MS 1512-cB58.  Pettini et
al.~(2000) estimate a mass loss rate of $\approx 60 \ {\rm M_\odot}$
yr$^{-1}$ and an outflow velocity of $V_\infty \approx 200$ km
s$^{-1}$.  The inferred star formation rate is $\dot M_\star \approx 40 \
{\rm M_\odot}$ yr$^{-1}$, suggesting a close correspondence between
the momentum input from stars and that in the outflow.


\subsection{The Faber-Jackson Relation}
\label{section:faber}

The Faber-Jackson (FJ) relation connects the luminosity of the bulge
or spheroidal component of a galaxy with its velocity dispersion
(Faber \& Jackson 1976; for a review, see Burstein et al.~1997).  Bernardi et al.~(2003) (their Fig.~4) give
the FJ relation derived from about 9,000 early-type galaxies in the
Sloan Digital Sky Survey.  In the i-band, their results imply 
\be 
\nu
L_{\nu,i} \simeq 2 \times 10^{44}\,\,{\rm
erg\,\,s^{-1}}\,\,\sigma_{200}^{3.95}
\label{fj}
\ee 
over the range $\sigma \approx 100-300 \ \kms$.  The slope of the FJ
relation is nearly identical in all of the Sloan bands, while the
normalization decreases slightly (by a factor of $\approx 2$) at the
shortest wavelengths (the $g$ band).  In a separate analysis Pahre et
al. (1998) give the FJ relation in the near-infrared (K-band), finding
$\nu L_{\nu,K} \simeq 3 \times 10^{43} \,\,{\rm
erg\,\,s^{-1}}\,\,\sigma_{200}^{4.1}$.  Thus the slope of the FJ
relation is essentially independent of wavelength while the change in
its normalization with wavelength is consistent with the spectrum of
an old stellar population.  That is, if one plots the normalization of
the FJ relation as a function of wavelength, the resulting
``spectrum'' is very similar to that produced by a $\sim 10$ Gyr old
instantaneous starburst.\footnote{Such an exercise is only possible
because the slope of FJ is independent of $\sigma$ so one can
meaningfully construct a ``spectrum.''  In particular, it is not
possible to carry out the same procedure for disk galaxies because the
slope of the Tully-Fisher relation increases with increasing
wavelength (Binney \& Merrifield 1998).}

Our expression for the limiting starburst luminosity is given in
eqs.~(\ref{lmom}) and (\ref{lmoms}).  This luminosity corresponds to
that {\it during} the starburst, whereas equation (\ref{fj}) is a statement
about $L$ and $\sigma$ {\it now}.  If, as we have argued above, most
(all) early type galaxies went through a significant starburst phase
during which their luminosities reached -- but did not exceed -- our
limiting luminosity, we can determine the properties of the stellar
population now by ``fading'' the starburst with time.

Figure \ref{plot:decay1} shows the luminosity of a starburst as a
function of time in the models of Bruzual \& Charlot (2003).
Starbursts of five durations are shown: instantaneous, 10 Myr, 30 Myr,
100 Myr, and 300 Myr (with constant star formation rates and a
Salpeter IMF between $0.1-100$ M$_\odot$).  The ratio of the peak
starburst luminosity to the luminosity now ($t\sim10^{10}$ yr)
determines how much the stellar population fades with time and allows
us to connect the maximum starburst luminosity to the currently
observed FJ relation.  For the models shown in Figure
\ref{plot:decay1} the starburst fades by a factor of $2500$, $1500$,
$800$, $250$, and $100$ over $\approx 10^{10}$ yrs.  These results can
be understood analytically by noting that, for a Salpeter IMF and a
stellar mass-luminosity relation of the form $L \propto M^\beta$, the
late-time luminosity of a starburst is given by \be L(t) \sim 3 L_{\rm
SB} \left({\tau_{\rm SB} \over \tau_{\rm max}}\right) \left({t \over
\tau_{\rm max}}\right)^{-{\beta - 1.35 \over \beta - 1}}
\label{fade} 
\ee where $\tau_{\rm max} \approx 3$ Myrs is the lifetime of the most
massive stars, $\tau_{\rm SB}$ (assumed $> \tau_{\rm max}$) is the
duration of the starburst, and $L_{SB}$ is the peak luminosity of the
starburst; the factor of ``3'' has been included based on comparison
to numerical calculations.  The dependence on $\tau_{\rm SB}/\tau_{\rm
max}$ seen in Figure \ref{plot:decay1} and equation (\ref{fade})
arises because the late-time luminosity is determined by {\it total}
number of low-mass stars made during the burst, while the peak
starburst luminosity ($L_{\rm SB}$) depends only on the {\it
instantaneous} number of massive stars present in the starburst.  For
$\beta \approx 4-5$, equation (\ref{fade}) predicts $L(t) \propto
t^{-0.9}$, in reasonable agreement with Figure \ref{plot:decay1} at
late times.

Comparing the observed FJ relation with the maximum starburst
luminosity in equation (\ref{lmom}) shows that if the starburst fades
by a factor of $\approx 100-200$ from $z \sim {\rm few}$ to now, then
we can account for both the normalization and slope of the FJ relation
as being due to feedback during the formation of ellipticals at high
redshift.  This in turn requires that most of the stars in a galaxy
were formed over a period of $\tau_{\rm SB} \sim 100-300$ Myrs
(Fig. 3).\footnote{The timescale $\tau_{\rm SB}$ refers to the net
timescale over which stars form at a luminosity $\sim L_{\rm M}$.  It
could in principle be that many starbursts of shorter duration
cumulatively last for $\sim \tau_{\rm SB}$.} This number is plausible
on a number of grounds.  It is comparable to the inferred star
formation timescales in LBGs (e.g., Shapley et al. 2001) and ULIRGs
(e.g., Genzel et al. 2004).  It is also comparable to the dynamical
timescale $\tau_{\rm Dyn}^{\rm D}$ of gas on galactic scales.  This is
relevant because this dynamical timescale roughly determines the
duration of starbursts in numerical simulations of merging galaxies
(Mihos \& Hernquist 1996).

The scatter in the FJ relation is observed to be a factor of $\approx
2$ in $L$ at a given $\sigma$ (Bernardi et al. 2003).  In our model,
this scatter is primarily due to differences in the time since, and
duration of, the star formation episode that built up most of the mass
of the galaxy.  Since most early type galaxies likely formed at $z
\sim 1-3$, and the time difference between these redshifts is only a
factor of $\approx 1.5$, the scatter produced in the observed FJ by
different ``formation redshifts'' is quite mild (since $L \sim
t^{-1}$; see Figure \ref{plot:decay1}).  By contrast, the amount by
which a starburst fades is directly proportional to its duration
$\tau_{\rm SB}$ (see eq.~[\ref{fade}]), which might {\it a priori} be
expected to vary significantly from system to system.  It is unclear
what would cause such a narrow range in $\tau_{\rm SB}$.  It is,
however, encouraging that the dynamical timescale at $\sim R_{\rm D}$
is independent of the mass ($\sigma$) of a galaxy
(eq.~[\ref{taudynd}]), suggesting that to first order the duration of
a merger-induced starburst might be similar in different
systems.\footnote{In numerical simulations of mergers, the duration of
a starburst depends on the details of the orbit and the internal
dynamics of the merger constituents (e.g., the bulge to disk ratio;
see Mihos \& Hernquist 1996).  Variations in these properties from
merger to merger will introduce scatter into $\tau_{\rm SB}$ and thus
the observed FJ relation.}  It is also possible, as we discuss in the
next section, that a central AGN is responsible for terminating the
star formation in its host galaxy.


\section{Active Galactic Nuclei \& the $M_{\rm BH}-\sigma$ Relation}
\label{section:agn}

Early-type galaxies and bulges are inferred to have central
supermassive black holes, whose masses correlate well with the
velocity dispersion of the galaxy itself: $M_{\rm BH} = 1.5 \times
10^8 \sigma_{200}^4 \ {\rm M}_\odot$ (Tremaine et al. 2002). This
correlation is remarkably similar to the FJ relation.

In \S\ref{section:momentum} we considered the general properties of galactic winds
driven by momentum deposition.  We then focused on radiation from
starbursts as providing this source of momentum.  However, star
formation is unlikely to efficiently remove gas from very small scales
in galactic nuclei (scales much smaller than that of a nuclear
starburst).  This gas is available to fuel a central AGN.

We consider a central BH with a luminosity $L_{\rm BH}$.  The
optically thin Eddington luminosity for the BH is (eq.~[\ref{lthin}])
\be 
L_{\rm Edd}={4\pi GM_{\rm BH}c\over\kappa_{\rm es}} = 
1.3\times10^{46}\,\,{\rm erg\,\,s^{-1}}\,\,M_8,  \label{edd}
\ee 
where $M_8=M_{\rm BH}/10^8$M$_\odot$ and $\kappa_{\rm es}=0.38{\rm\ cm^2/g}$ is the
electron scattering opacity.  Note that the electron scattering
opacity is appropriate close to the BH, at least out to the dust
sublimation radius. The latter can be estimated by equating the
absorbed flux with the radiated flux from dust grains: 
\beqa 
\label{eq:rsub}
R_{\rm Sub}&=&\sqrt{L_{\rm BH}\over 4\pi\sigma_{SB} T^4_{\rm Sub}}
\nonumber \\
&\sim&3\times10^{18}L^{1/2}_{46}\left(1200{\rm \,\,K}\over T_{\rm
Sub}\right)^2{\rm cm}, 
\eeqa 
where $L_{46}=L_{\rm BH}/10^{46}$ erg
s$^{-1}$, $\sigma_{SB}$ is the Stefan-Boltzman constant, and the dust
sublimation temperature is $T_{\rm Sub}\approx 1200$ K.

The ratio $\Gamma\equiv L_{\rm BH}/L_{\rm Edd}$ is estimated to be
$\sim 0.1-1$ for luminous AGN at high redshift (e.g., Vestergaard
2004). If the hole radiates with an efficiency $\eta\approx0.1$, the
mass accretion rate is $\dot{M}_{\rm BH}=L_{\rm BH}/(\eta c^2)$.
Combining $\dot{M}_{\rm BH}$ and $L_{\rm Edd}$ gives the timescale for
$L_{\rm BH}$ (and $M_{\rm BH}$) to double, the Salpeter timescale, 
\be
\tau_{\rm Salp}=\frac{\eta c\kappa_{\rm es}}{\Gamma4\pi G}\sim 43\
{\rm Myrs}\,\,\,\Gamma^{-1}.
\label{tsalp}
\ee 
The region exterior to the sublimation radius contains dust and
can be optically thick to the UV photons of the AGN even if the AGN is
sub-Eddington in the electron scattering sense (eq. [\ref{edd}]).
This is simply because the dust opacity is much larger than the
electron scattering opacity.  Thus, by arguments analogous to those
given in \S\ref{section:momentum} and \S\ref{section:starburst}, if
the luminosity of the black hole exceeds $L_{\rm M}$
(eq.~[\ref{lmom}]), it drives an outflow.  This outflow drives away
gas outside of $R_{\rm Sub}$, irrespective of whether or not the AGN
is super-Eddington on small scales close to the BH.

Using $L_{\rm BH}=\Gamma L_{\rm Edd}$ the criterion $L_{\rm BH}
\approx L_{\rm M}$ can be written in terms of the black hole mass as
\be 
M_{\rm BH} \approx {f_g\kappa_{es}\over \pi G^2
\Gamma}\sigma^4\approx2\times10^8\,{\rm
M_\odot}\,\,\fgas\Gamma^{-1}\sigma_{200}^4.  
\label{msigma} 
\ee 
If the
black hole mass exceeds the limit in equation (\ref{msigma}), then it
drives a large-scale galactic outflow.  Only when $M_{\rm BH}$ reaches
the critical mass in equation (\ref{msigma}) will it be able to blow dusty
gas all the way out of the galaxy.  This shuts off the gas supply to
the black hole on a dynamical timescale and fixes the mass to be that
in equation (\ref{msigma}), in good agreement with the observed $M_{\rm
BH}-\sigma$ relation.  It should be noted that the dust-free gas
within $R_{\rm Sub}$ need not be blown out by the BH.  The total mass
contained within this region is, however, a small fraction ($\sim$ few
\%) of the BH mass (eq.~[\ref{msigma}]), so accretion of this gas does
not modify the $M_{\rm BH}-\sigma$ relation.

Although the context is somewhat different, equation (\ref{msigma}) is
identical to the $M_{\rm BH}-\sigma$ relation derived by King (2003) and it is similar to 
those obtained using other `feedback' arguments for the $M_{\rm BH}-\sigma$ relation 
 (e.g., Silk \& Rees 1998; Haehnelt et al. 1998; Blandford 1999; Fabian 1999; Fabian et al.~2002).  
Specifically, King
assumed that a radiation pressure driven outflow launched from close
to the BH sweeps out of the galaxy, driving all of the gas away.  We
argue that the outflow is primarily due to absorption of the BHs
luminosity by dust outside of $R_{\rm Sub}$, independent of whether or
not the AGN drives an outflow from small radii $\ll R_{\rm Sub}$.

An interesting feature of our model -- or, more generally, of
observations of AGN and starbursts -- is the apparent coincidence that
the Salpeter time that governs the growth of the BH is comparable to
the duration of the star formation epoch (see
\S\ref{section:starburst} for a discussion of the latter).  Were the
Salpeter time much shorter, the BH would grow rapidly and its outflows
could significantly disrupt star formation before sufficient stars
formed to lie on the FJ relation.  As is, we suggest that {both} the
star formation and BH growth are independently self-regulating,
reaching the maximum luminosity $\sim L_{\rm M}$ (eq.~[\ref{lmom}]).
However, as explained in the previous section, it is unclear what
determines the duration of the star formation epoch.  This may be
determined by mergers, but it is also possible that the `coincidence'
between $\tau_{\rm Salp}$ and $\tau_{\rm SB}$ is no coincidence at
all: when the BH reaches the mass given in equation (\ref{msigma}) it
drives an outflow that sweeps out from the galactic nucleus,
terminating star formation in its host galaxy (e.g., Silk \& Rees
1998; Fabian 1999).  This possibility is interesting because the
Salpeter time is likely to be similar in different systems, which
could explain the narrow range of $\tau_{\rm SB}$ required to
understand the FJ relation.

The above discussion assumes that BHs reach, but do not significantly
exceed, the luminosity $L_{\rm M}$.  In Figure \ref{plot:lsagn} we
test this prediction using data compiled by Boroson (2003) and Shields
et al.~(2003).  Both papers estimate the velocity dispersion of
galaxies hosting quasars using the width of the narrow OIII line (see
Nelson 2000).  The bolometric luminosity is estimated using $L \approx
9 \nu L_\nu(5100 \, \AA)$, the average bolometric correction used by
Kaspi et al. (2000).  There is evidence that the width of the OIII
line can sometimes exceed the velocity dispersion of the galaxy in
radio-loud AGN (Nelson \& Whittle 1996); these systems are indicated
by open symbols in Figure \ref{plot:lsagn}.

Figure \ref{plot:lsagn} shows that the limit $L_{\rm M}$ accounts for
the maximum quasar luminosity at any $\sigma$, in good agreement with
the predictions of feedback models for the $M_{\rm BH}-\sigma$
relation.  That some systems lie below $L_{\rm M}$ is not surprising
because most BHs spend most of their time accreting at sub-Eddington
rates; note also that Boroson's sample from SDSS contains only quasars
with $z \lesssim 0.5$ and thus systematically lacks high redshift,
high luminosity quasars.


\section{Discussion}
\label{section:conclude}

\subsection{Galactic Winds}

In this paper, we have investigated large-scale galactic winds driven
by momentum deposition, in contrast to the usual assumption that
energy deposition (thermal heating) by core-collapse supernovae drives
these outflows.  The efficiency of energy-driven outflows is uncertain
because much of the energy deposited by supernovae in the ISM may be
radiated away.  Even in this limit, momentum injection by supernovae
is important and can itself generate a powerful outflow.  Supernovae
contribute to `momentum-driving' in a second way: the dynamics of cold
gas entrained in a hot flow is analogous to that of a momentum-driven
wind (see \S\ref{section:entrainment} and the Appendix).  Note that
these mechanisms are physically distinct.  The latter (ram pressure
driving of cold gas) requires a powerful hot wind, while the former
operates even if the supernovae energy is radiated away.

In addition to supernovae, momentum injection is provided by continuum
absorption and scattering of radiation on dust grains (radiation
pressure); such radiation can be produced by either a starburst or a
central AGN (or both) and is an efficient mechanism for driving cold,
dusty gas out of a galaxy.
Interestingly, the forces due to radiation pressure and ram pressure
(entrainment) may be comparable in many cases (see
eq.~[\ref{ramvsrad}]).  Distinguishing which mechanism dominates is
non-trivial.  One way may be to assess the mass loss rate in hot gas
via X-ray observations (see, however, Strickland \& Stevens 2000 who
argue that such observations don't necessarily probe the
energy-containing phase of the hot wind).

Although uncertain, we suggest that momentum injection may be more
effective at halting star formation and `blowing away' the gas in a
galaxy than energy injection.  For example, supernovae energy can be
efficiently vented by `blowing out' of the galactic disk, even if
little of the mass is lost (De Young \& Heckman 1994).  By contrast,
the momentum of supernova explosions cannot be similarly vented and
thus may be more disruptive to the bulk of the gas in a galaxy.  In
addition, because the mass of a galaxy is primarily in the cold phase,
radiation pressure and ram pressure driving of cold gas may dominate
the mass loss in starbursting galaxies (even in the presence of a hot
thermal wind).  

Momentum-driven winds have several properties that may allow them to
be distinguished from energy-driven winds
(\S\ref{section:momentum}). Specifically, (1) the momentum flux in the
outflow, $\dot M_{\rm W} V_\infty$, is comparable to that in the
radiation field, $L/c$ (eq.~[\ref{mom}]) and (2) the terminal velocity
of the outflow should be comparable to the velocity dispersion of the
host galaxy, $V_\infty \sim \sigma$ (eqs.~[\ref{vthick}] \&
~[\ref{vthin}]). Note that these predictions apply to outflowing cold
gas driven by momentum-deposition.  The hot thermally-driven phase of
a galactic wind satisfies different scalings (see
\S\ref{section:energy}).

The simple predictions above for mass loss rates and terminal
velocities could be readily incorporated into cosmological simulations
to assess the global impact of momentum-driven galactic winds (as in
the work of Aguirre et al. 2001a).  One interesting possibility is
that because the energy carried by a momentum-driven wind may be
smaller than that of a thermal supernovae-driven wind\footnote{The
energy carried by a momentum-driven wind is $\sim V_\infty L/c \sim
\sigma L/c$ while that in a thermal supernovae-driven wind is $\sim
10^{-2} \xi L$.}, momentum-driven winds may pollute the intergalactic
medium with metals without significantly modifying its structure from
that predicted by the gravitational instability paradigm.

\subsection{The Growth of Ellipticals and Black Holes}

In addition to considering the general properties of momentum-driven
galactic winds, we have derived a limiting luminosity, $L_{\rm
M}\simeq(4f_g c/G)\,\sigma^4$, above which momentum-deposition is
sufficient to drive away a significant fraction of the gas in a galaxy
(eq.~[\ref{lmom}]; King 2003 derived a similar result in the context of
black hole growth; Meurer et al.~1997 and Lehnert \& Heckman 1999 discuss
a potentially related observational limit on the surface brightness of local and 
high-$z$ starburst galaxies).  This outflow may regulate star formation during the
formation of ellipticals at high redshift by limiting the gas available for star formation, 
ensuring that the luminosity never significantly exceeds $L_{\rm M}$.
The fact that massive starbursts have
luminosities near $L_{\rm M}$ (Fig.~\ref{plot:ls}; \S\ref{section:starburst}) and that 
starbursts at $z \gtrsim 1$ account for a significant fraction of the
local stellar inventory (e.g., Madau et al. 1998) supports a
model in which, during the hierarchical growth of galaxies, mergers
trigger intense starbursts ($L \sim L_{\rm M}$) that form a
significant fraction of the stars in early-type galaxies.

We have focused on the growth of elliptical galaxies, rather than
spirals, because there is evidence that star formation in spirals is
reasonably quiescent (e.g., Kennicutt et al. 1994) and it is thus
unlikely that a significant fraction of the mass in spiral galaxies
was formed during `bursts' that reached our limiting luminosity $L_{\rm M}$.  By
contrast, ellipticals are inferred to have formed most of their stars
relatively quickly at high redshift $\gtrsim 1-3$ (e.g., Van Dokkum et
al. 2004).

Our hypothesis that proto-elliptical galaxies at high redshift go
through an extended period of star formation with $L \sim L_{\rm M}
\propto \sigma^4$ can explain the Faber-Jackson relation between the
{\it current} luminosity and velocity dispersion of elliptical
galaxies (\S\ref{section:faber}).  Specifically, our model explains
quantitatively why ellipticals do not have $L \propto M_{\rm DM}
\propto \sigma^3$ (where $M_{\rm DM}$ is the total mass of the dark
matter halo), as would be expected if a fixed fraction of the
available gas were converted into stars.  Our model also explains why elliptical
galaxies do not have $L \propto \sigma^5$, which would be expected if
energy-deposition from supernovae dominated feedback (Fig.~\ref{plot:lss}).
Since the luminosity of a starburst is dominated by the rate at which
high mass stars are being formed, while the current luminosity of
ellipticals (reflected in FJ) depends on the total number of low-mass
stars in the galaxy, our interpretation of FJ requires that the
duration of peak star formation activity was relatively similar in
different galaxies (so that both the peak star formation rate and the
total number of stars formed are similar).\footnote{Note that this
will be a requirement for any model that tries to explain the FJ
relation as a result of feedback during starbursts because `feedback'
(generically defined) is sensitive to the star formation rate, rather
than the total number of stars formed.}  By comparing the limiting
luminosity $L_{\rm M}$ with the current Faber-Jackson relation, we
infer a star formation duration of $\sim 100-300$ Myrs (\S\ref{section:faber}).  This
is in reasonable agreement with observational inferences in LBGs
(e.g., Shapley et al. 2001) and ULIRGs (e.g., Genzel et al. 2004).

Our model for the origin of the FJ relation does not fully explain why
ellipticals lie in the fundamental plane.  Roughly speaking the
fundamental plane can be understood via two projections: $L\propto\sigma^4$ (the FJ
relation) and $R_{\rm eff}\propto\sigma^{8/3}$, where $R_{\rm eff}$ is the effective radius (Bernardi et al. 2003).  
The latter relation is very different from any virial prediction, which would suggest $R_{\rm eff} \propto \sigma$,
and its origin, whether a consequence of gas physics or collisionless mergers of stellar
systems in unknown.

Unlike ellipticals, the optical Tully-Fisher relation in spirals is reasonably consistent with $L \propto
v_c^3$ (where $v_c$ is the maximum circular velocity; e.g., Giovanelli
et al.~1997).  This is probably a consequence of the more quiescent
star formation histories of spirals (Kennicutt et al.~1994) so that `feedback' is less severe
and the luminosity of a galaxy is simply proportional to its mass.
However, the slope of the TF relation varies systematically with
wavelength and in the IR, $L \propto v_c^4$ (e.g., Pierini \& Tuffs
1999), consistent with the FJ scaling for ellipticals.  This is very
intriguing and might suggest that the oldest stars in spirals were
formed in bursts analogous to those that formed ellipticals.

\subsubsection{Black Holes}

In addition to considering the self-regulated growth of elliptical
galaxies via starbursts, we propose that the growth of black holes in
early type galaxies proceeds in a similar manner.  As a black hole
grows via accretion, its luminosity may eventually exceed $\sim L_{\rm
M}$.  When it does so, the dusty
gas around the black hole (outside the sublimation radius;
eq.~[\ref{eq:rsub}]) is blown away by radiation pressure.  The black
hole thus shuts off its own fuel supply. This fixes the BH mass to lie
very close to the observationally inferred $M_{\rm BH}-\sigma$
relation (see eq.~[\ref{msigma}]).  If star formation in the host
galaxy is still ongoing when the BH reaches $\sim L_{\rm M}$, the
outflow from the galactic nucleus may sweep through the galaxy,
terminating star formation.  This possibility is interesting because
it may explain the apparent coincidence that the Salpeter time
characterizing the growth of black holes (eq.~[\ref{tsalp}]) is
similar to the inferred duration of star formation in high redshift
starbursts (see \S\ref{section:starburst} and \S\ref{section:agn}).

Previous discussions of the interaction between a central black hole
and its surrounding galaxy have also emphasized how the central black
hole can regulate its own fuel supply by driving away ambient gas
(e.g., Silk \& Rees 1998; Haehnelt et al. 1998; Blandford 1999; Fabian
1999; King 2003).  All such models are broadly similar (ours
included), though they differ in detail as to whether energy
deposition or momentum deposition is the most important feedback
mechanism.  
However, implicit in previous discussions of the $M_{\rm
BH}-\sigma$ relation is that the stars in the galaxy ``know'' when the
hole is about to reach the limiting mass at which it can blow away the
surrounding gas.  Otherwise it is unclear how the right number of
stars are formed so that the galaxy lies on the FJ relation.  One
explanation for this is to hypothesize that the stars form as the gas
is being blown out by the AGN, i.e., in one dynamical time (e.g., King
2003). Observationally, however, this is not the case in either LBGs
or ULIRGs, where the star formation lasts for 100s of Myrs.  Instead,
we argue that the AGN's role may be sub-dominant: feedback from stars
determines the maximal luminosity of a starburst, whether or not there
is an AGN present.  It is, however, possible that the AGN administers
the coup de gr$\hat{\rm a}$ce, terminating star formation.

In our interpretation, the peak episode of star formation likely
precedes that of AGN activity in most galaxies. There are two reasons 
for this.  First, gas is transported from the outside in.  Therefore, star formation on galactic scales sets in
before the central BH is fed.  Second, if the BH were to grow and reach
the $M_{\rm BH}-\sigma$ relation {\it before} significant star formation has
occurred, it will blow out the ambient gas in the galaxy before the FJ relation is set.
If the galaxy were to later accrete
gas -- from the IGM or via a merger -- because the BH is already on
$M_{\rm BH}-\sigma$ relation even a small amount of accretion onto the
BH would be sufficient to again disrupt star formation.
There is some observational support for this temporal ordering.  First, the
number density of bright quasars declines more rapidly at high $z$
than the number density of star forming galaxies (compare Fan et
al. 2004 and Heavens et al. 2004).  Second, although some rapidly star-forming 
SCUBA sources at high $z$ are inferred to host quasars, in
many cases there is X-ray evidence for more modest AGN with $L \sim
10^{43}-10^{44}$ ergs s$^{-1}$ (e.g., Alexander et al. 2003).  Since
many of the observed systems are Compton thin, it is unlikely that a
quasar-like luminosity is hidden by obscuration.  Given the inferred
$\sigma \sim 200-300 \kms$ in the SCUBA sources (\S\ref{section:starburst} and Table \ref{tab:ls}), BHs
on the $M_{\rm BH}-\sigma$ relation would have $M \sim 10^8-10^9
M_\odot$.  To explain the observed luminosities would then require
substantially sub-Eddington accretion rates.  While possible, this
would be surprising in view of the large available gas supply.  It is perhaps
more plausible that the BH is still growing and has not yet reached
the $M_{\rm BH}-\sigma$ relation (e.g., Archibald et al. 2002).

Our model makes the very strong prediction that the peak luminosity of
star formation and AGN activity in a given galaxy are essentially the
same ($\sim L_{\rm M}$), set by the criterion that a momentum-driven
outflow blows away gas that would otherwise be available for star
formation/accretion.  Figures \ref{plot:ls} and \ref{plot:lsagn}
provide observational evidence that is consistent with this
prediction.  
%
We note, however, that in order to reach a luminosity $\sim L_{\rm M}$, a galaxy with $\sigma
= 200 \, \sigma_{200}$ km s$^{-1}$ must have a star formation rate of
$\approx 500 \, \sigma_{200}^4\, {\rm M_\odot \, yr^{-1}}$
(eq.~[\ref{mdotstar}]), while a black hole must accrete gas at
$\approx 5 \, \sigma_{200}^4 \, {\rm M_\odot \, yr^{-1}}$.  Moreover,
if the most luminous observed AGN ($L \sim 10^{48}$ ergs s$^{-1}$, see
Fig. \ref{plot:lsagn}) are Eddington-limited and lie on the $M_{\rm BH}-\sigma$
relation, then their host galaxies must have $\sigma \sim 500 \kms$. If
such galaxies indeed exist and if star formation is to reach a
luminosity $\sim L_{\rm M}$, then the required star formation rate is $\sim
10^4 \, {\rm M_\odot \, yr^{-1}}$!  Such a starburst has never been
observed, but would of course be extremely rare.  It is unclear
whether such a star formation rate can actually be achieved and sustained.  If not,
then we predict deviations from FJ for the largest ellipticals
($\sigma>\sigma_{\rm max}$; see Fig. 1).  

\acknowledgments

We thank Crystal Martin for an inspiring talk that motivated this
work, and for useful conversations.  We also thank Alice Shapley, Leo
Blitz, Martin White, James Graham, Avishai Dekel, Reinhard Genzel,
Anthony Aguirre, Chung-Pei Ma, Yoram Lithwick, Volker Springel, and
Jon Arons for helpful conversations.  We thank A. Bruzual, S. Charlot,
and C. Leitherer for making their starburst models available and 
Gabriela Mall\'en-Ornelas for providing us with data from her thesis.
We thank the referee for useful comments.
N.M. is
supported in part by the Canada Research Chair program and the Miller
Foundation.  N.M. extends his gratitude to UC Berkeley where much of
this work was completed.  T.A.T. is supported by NASA through Hubble
Fellowship grant \#HST-HF-01157.01-A awarded by the Space Telescope
Science Institute, which is operated by the Association of
Universities for Research in Astronomy, Inc., for NASA, under contract
NAS 5-26555.  E.Q. is supported in part by NSF grant AST 0206006, NASA
grant NAG5-12043, an Alfred P. Sloan Fellowship, and the David and
Lucile Packard Foundation.


\begin{appendix}

\section{Energy-Driven Galactic Winds: Entrainment}

In this Appendix, we consider entrainment in two limits: (1)
entrainment of individual clouds (the ``optically thin'' limit) and
(2) entrainment of shells of gas (the ``optically thick'' limit,
appropriate when the cold gas occupies a large fraction of $4 \pi$ sr
on the sky and thus intercepts much of the momentum flux in the hot
flow).  We show that in both limits the dynamics of cold gas entrained
in a hot flow is analogous to that of the momentum-driven winds
considered in \S\ref{section:momentum}.

\subsection{Entrainment:  The `Optically Thin' Limit} 

\label{section:entrainthin}

We first consider the entrainment of individual clouds of cold gas;
the clouds have projected area $A_c$, density $\rho_c$, and mass $M_c
= (4\pi/3) \rho_c R_c^3 = (4/3) A_c R_c \rho_c$.  The hot wind has a
density $\rho_h = \dot M_h/(4\pi r^2 V_h)$, a mass loss rate $\dot
M_h$, and a velocity $V_h$.\footnote{In this section, we use the
subscript `h' for `hot' -- rather than than our previous subscript `W'
for `wind' -- to emphasize that these mass loss rates and velocities
refer to that of the hot flow.}  The ram pressure force on a cold
cloud is $\rho_h V_h^2 A_c$.  Comparing this to the radiation pressure
force on the cloud (assuming it is optically thick to radiation)
yields
\beqa 
\frac{F_{\rm ram}}{F_{\rm rad}} = \frac{\rho_h V_h^2
A_c}{(L/c)(A_c/4\pi r^2)} & = & \left(\frac{\dot{M}_h}{\dot{M}_\star
}\right)\left(\frac{V_h}{300 \ \epsilon_3 \ {\rm km \ s^{-1}}}\right)
\nonumber \\ & \approx & \left({600\,\xi_{0.1}\,\kms \over
V_h}\right),
\label{ramvsrad}
\eeqa 
where $A_c/(4\pi r^2)$ is the fraction of photons intercepted by the
cloud and in the last equality we used $\dot E_{\rm SN} = {1 \over 2}
\dot M_h V_h^2 = 10^{-2} \xi L$.  Equation (\ref{ramvsrad}) shows that
ram pressure and radiation pressure can contribute comparably to the
driving of cold gas, so long as $\dot M_h\sim \dot M_\star$.  Shocked
supernova ejecta contribute a total mass loss of $\sim 0.1 \dot
M_\star$ in the absence of radiative cooling.  It is, however,
plausible that considerable swept up mass is also shock heated,
leading to $\dot M_h \sim \dot M_\star$ (see, e.g., Martin 1999 for
evidence to this effect in local starbursts).  Note that $V_h \sim
300-600 \kms$ is consistent with the observed temperature of hot
outflowing gas in local starbursts (e.g., Martin 1999).

In the limit that ram pressure dominates the driving of cold gas, we
can derive the velocity of the cloud as a function of distance from
the galaxy by analogy with the optically thin radiation pressure limit
considered in \S\ref{section:momentum} (since $F_{\rm ram} = \rho_h
V_h^2 A_c \propto r^{-2}$, the optically thin limit, rather than the
optically thick limit, is the appropriate analogy).  The velocity
profile is given by \be V(r)=\sqrt{V_c^2 \left({1 - {R_0 \over
r}}\right) - 4 \sigma^2\ln(r/R_0)},
\label{vram}
\ee where $R_0$ is the initial `launching' radius and \be V_c^2 = {3
{\dot M_h} V_h \over 8 \pi \rho_c R_c R_0}. \label {vc} \ee The
velocity $V_c$ is the characteristic velocity the cloud reaches before
it begins to decelerate in the extended gravitational potential of the
galaxy.  Note that equation (\ref{vc}) is only appropriate for $V_c < V_h$;
if equation (\ref{vc}) predicts $V_c > V_h$, the actual maximal velocity is
$V_h$ since ram pressure ceases to accelerate the cloud above this velocity.

In order for the cloud to move to a radius significantly larger than
its starting position at $\sim R_0$, we require $V_c \gtrsim 2
\sigma$.  Using $\rho_c R_c = m_p N_H$, this requirement can be
rewritten as \beqa \dot M_h & \gtrsim & {32 \pi \sigma^2 m_p N_H R_0
\over 3 V_h} \nonumber \\ & \approx & 35 \, {\rm M_\odot \,yr^{-1}} \,
\sigma_{200}^2 \, N_{21} \left(R_0 \over 1 \,{\rm kpc}\right)\left(300
\,{\rm km/s} \over V_h\right) \label{mdothot} \eeqa If
$\dot{M}_h$ is less than the value given in equation (\ref{mdothot}),
cold clouds cannot be pushed out of the nuclear region by the hot
flow.  This criterion is analogous to the optically thin Eddington
limit given in equation (\ref{leddcloud}).  Indeed, if $\dot M_h
\approx \dot M_\star$, then $F_{\rm ram} \approx F_{\rm rad}$
(eq. [\ref{ramvsrad}]) and so the two `Eddington-limits' are
essentially equivalent.


The cloud velocity $V_c$ can be rewritten as
\beqa 
{V_c \over V_h} &=& \sqrt{3 {\dot M_h} \over 8 \pi m_p N_H V_h
R_0} \label{vcscale} \\ & \approx & \left[\left({\dot M_h} \over
{\rm 10\,M_\odot/yr}\right) \left(10^{21} \, {\rm cm^2} \over N_H
\right) \left(1 \, {\rm kpc} \over R_0 \right) \left(300 \, {\rm km/s}
\over V_h \right)\right]^{1/2} \nonumber 
\eeqa 
Equation (\ref{vcscale}) implies that, for clouds of a given column $N_H$,
there is a critical mass loss rate $\sim 10 \mpy$ (in the hot phase)
below which ram pressure is {insufficient} to accelerate clouds of
cold material to a velocity $\sim V_h$, i.e., to the velocity of the
hot wind.  Assuming that $\dot M_h \sim \dot M_\star$, this can be
equivalently interpreted as defining a critical star formation rate
$\dot M_c \sim 10 \mpy$ such that for $\dot M_\star \gtrsim \dot
M_c$, $V_c \approx V_h$.  However, if $\dot M_\star \lesssim \dot M_c$
then $V_c \lesssim V_h$ with $V_c \propto \sqrt{{\dot M_\star}/[N_H
R_0]}$.  Note that the precise value of $\dot M_c$ is uncertain
because it depends on the column $N_H$ and the launching radius $R_0$.

\subsection{Entrainment:  The `Optically Thick' Limit} 
\label{section:entrainthick}

If the cold gas covers a significant fraction of the sky, it will
intercept most of the momentum flux in the hot flow.  The ram pressure
force on the cold gas is then given by $\rho_h v^2_h 4 \pi r^2$,
rather than $\rho_h v^2_h A_c$ as considered in the previous section.
We call this the `optically thick' limit.  In this limit, the
requirement to blow out all of the cold gas in the galaxy by ram
pressure is that
\be \rho_h v_h^2 4 \pi r^2 \gtrsim {G M M_g \over r^2} \approx {4 f_g
\sigma^4 \over G},
\label{cloudthick} 
\ee 
where we assume that most of the mass $M_g$ is in the cold phase.  Equation
(\ref{cloudthick}) can be rewritten as a requirement on the luminosity
of the starburst as follows: we rewrite $\rho_h$ in terms of $\dot
M_h$ and then use $\dot E_{\rm SN} = {1 \over 2} \dot M_h v_h^2 =
10^{-2} \xi L$ to find \be L
\gtrsim L^{\rm ent}_{\rm M} \equiv \frac{4\,f_g\,c}{G}\,\,\sigma^4 \,
\left({V_h \over 600\,\xi_{0.1}\,\kms}\right). \label{lmoment} \ee
Equation (\ref{lmoment}) is essentially the same as equation
(\ref{lmom}) -- the criterion to blow away all of the gas via a
momentum-driven wind.
Note, however, that the normalization in equation (\ref{lmoment})
depends on $\xi$, the efficiency with which SN energy is thermalized
in the ISM.  For our fiducial value of $\xi \approx 0.1$, $L^{\rm
ent}_{\rm M} \approx L_{\rm M}$, i.e., the limits set by ram pressure
and radiation pressure driving are comparable.  By contrast if $\xi
\sim 1$ ram pressure dominates, while if $\xi \lesssim 0.1$, radiation
pressure dominates.

\end{appendix}

\figcaption{Schematic diagram of the limiting starburst luminosity for
momentum-driven galactic winds as a function of $\sigma$ ($L_{\rm
M}\propto\sigma^4$; eq.~[\ref{lmom}]; thick solid line).  The thin solid
lines show the maximum attainable starburst luminosity determined by
converting all of the available gas into stars on a dynamical
time ($L_{\rm max}\propto\sigma^3$; eq.~[\ref{lmax}]) and the limiting
starburst luminosity for energy-driven galactic winds ($L_{\rm
E}\propto\sigma^5$; eq.~[\ref{lmom}]).  For $\sigma<\sigma_{\rm crit}$
(eq.~[\ref{sigmacrit}]), energy-driven winds dominate wind driving and
feedback because $L_{\rm E}<L_{\rm M}$.  The value of $\sigma_{\rm
crit}$ depends sensitively on the fraction of the kinetic energy
injected by SN that is radiated away (\S\ref{section:compare}).  Because the starburst
luminosity is bounded by $L_{\rm max}$, for $\sigma>\sigma_{\rm max}$
(eq.~[\ref{sigmacritd}]), the starburst cannot reach $L_{\rm M}$.  In
the intermediate region where $\sigma_{\rm crit}<\sigma<\sigma_{\rm
max}$, the starburst luminosity is bounded by $L_{\rm M}$ and
momentum-deposition dominates wind driving and feedback.  As we argue
in \S\ref{section:starburst}, this effect sets the Faber-Jackson
relation for elliptical galaxies.  The FJ relation, $L_{\rm
FJ}\propto\sigma^4$ is sketched here for comparison with the other
luminosity limits (thick dashed line).  It has a lower "zero-point"
than $L_{\rm M}$ as a result of passive, post-starburst evolution of
the stellar population.  This schematic plot is to be compared with
Fig.~\ref{plot:ls}, which shows the observed FJ relation and data from
local and high redshift starburst galaxies.  Figure 2 suggest that the
limit $L_{\rm M}$ -- rather than $L_{\rm E}$ -- is relevant down to
quite small $\sigma \sim 20 \, \kms$.\label{plot:lss}}

\figcaption{The limiting luminosity for momentum-driven galactic winds
($L_{\rm M}$; eq.~[\ref{lmom}]) as a function of the velocity
dispersion $\sigma$, for three values of the gas fraction ($f_g=0.1$,
thick solid line; $f_g=1.0$ and 0.01, dashed lines).  These curves
also account for uncertainty associated with the net momentum
deposition rate in starbursts, which includes contributions from
radiation, stellar winds, and supernovae.  Also shown is the observed
Faber-Jackson relation from eq.~(\ref{fj}).  From
\S\ref{section:momentum} and \S\ref{section:starburst}, we predict
that for $\sigma_{\rm crit}<\sigma<\sigma_{\rm max}$ (see
Fig.~\ref{plot:lss}), no system should have a luminosity greater than
$L_{\rm M}$.  We test this prediction in this figure by surveying the
literature for the brightest objects at any $\sigma$.  Detailed
information on all systems plotted here can be found in Tables
\ref{tab:ls} and \ref{tab:ls3} and the references cited.  The open
squares show high redshift ULIRGs ($z\sim2-3$), taken from Genzel et
al.~(2003), Neri et al.~(2003), and Tecza et al.~(2004).  The solid
squares show local ULIRGs (Genzel et al.~2001; Tables 1 and 2).  The
sample of $z\sim3$ LBGs are the open circles (Pettini et al.~2001)
while the $z\sim2$ LBGs are the ``X''s (Erb et al.~2003).  The open
triangles are taken from the sample of blue CFRS galaxies at
$z\sim0.6$ in Mall\'{e}n-Ornelas et al.~(1999) (their Fig.~2).  We
have selected the brightest galaxies at several $\sigma$.  The filled
triangles show the sample of HII galaxies from Telles \& Terlevich
(1997) (their Tables 1 and 3).  The open stars are dwarf galaxies from
Martin (1998) and Mateo (1998) (Tables 4 and 7).  The subset of dwarfs
plotted here are those with the highest luminosities associated with
current star formation, rather than the old stellar
population. Finally, we include a selection of local starbursts from
Heckman et al.~(2000) that show evidence for outflows of cold gas (see
Table \ref{tab:ls3} for details).
\label{plot:ls}}

\figcaption{Starburst luminosity as a function of time in the models of
Bruzual \& Charlot (2003), for an instantaneous starburst, and
starbursts with duration $\tau_{\rm SB}=$10, 30, 100, and 300 Myr.
All calculations were normalized to the same peak luminosity and
employ a Salpeter IMF from $0.1-100$ M$_\odot$.  We argue that the
ratio between the peak starburst luminosity and the luminosity at
$10^{10}$ yr gives the normalization between $L_{\rm M}$
(eqs.~\ref{lmom} and \ref{lmoms}) and the present-day Faber-Jackson
relation (eq.~[\ref{fj}]).  This requires a starburst lifetime of
$\tau_{\rm SB}\sim300$ Myr.
\label{plot:decay1}}

\figcaption{Same as Fig.~\ref{plot:ls}, but for quasars from the SDSS
sample of Boroson (2003) (squares) and the compilation of Shields et
al.~(2003) (circles).  The velocity dispersion is estimated from the
OIII linewidth while the luminosity is estimated using $L \approx 9
\nu L_\nu(5100 \, \AA)$, the bolometric correction advocated by Kaspi
et al. (2000).  The limiting luminosity for momentum-driven galactic
winds ($L_{\rm M}$; eq.~[\ref{lmom}]) is also shown, for three values
of the gas fraction ($f_g=0.1$, thick solid line; $f_g=1.0$ and 0.01,
dashed lines).  This limit accounts reasonably well for the maximum
quasar luminosity at any $\sigma$.  Note that Boroson (2003) does not
present his observed values of $\nu L_\nu(5100 \, \AA)$, but they can
be determined from his inferred black hole masses and $H-\beta$
linewidths ($v_{H\beta}$) using $M_{\rm BH} = 3v_{H\beta}^2 R_{\rm
BLR}/(4G)$ and $R_{\rm BLR} = 34 [\nu L_\nu(5100 \, \AA)/10^{44} \,
{\rm ergs \, s^{-1}}]^{0.7} \, {\rm lt-days}$, where $R_{\rm BLR}$ is
the radius of the broad line region (see his Table 1).
\label{plot:lsagn}}

\plotone{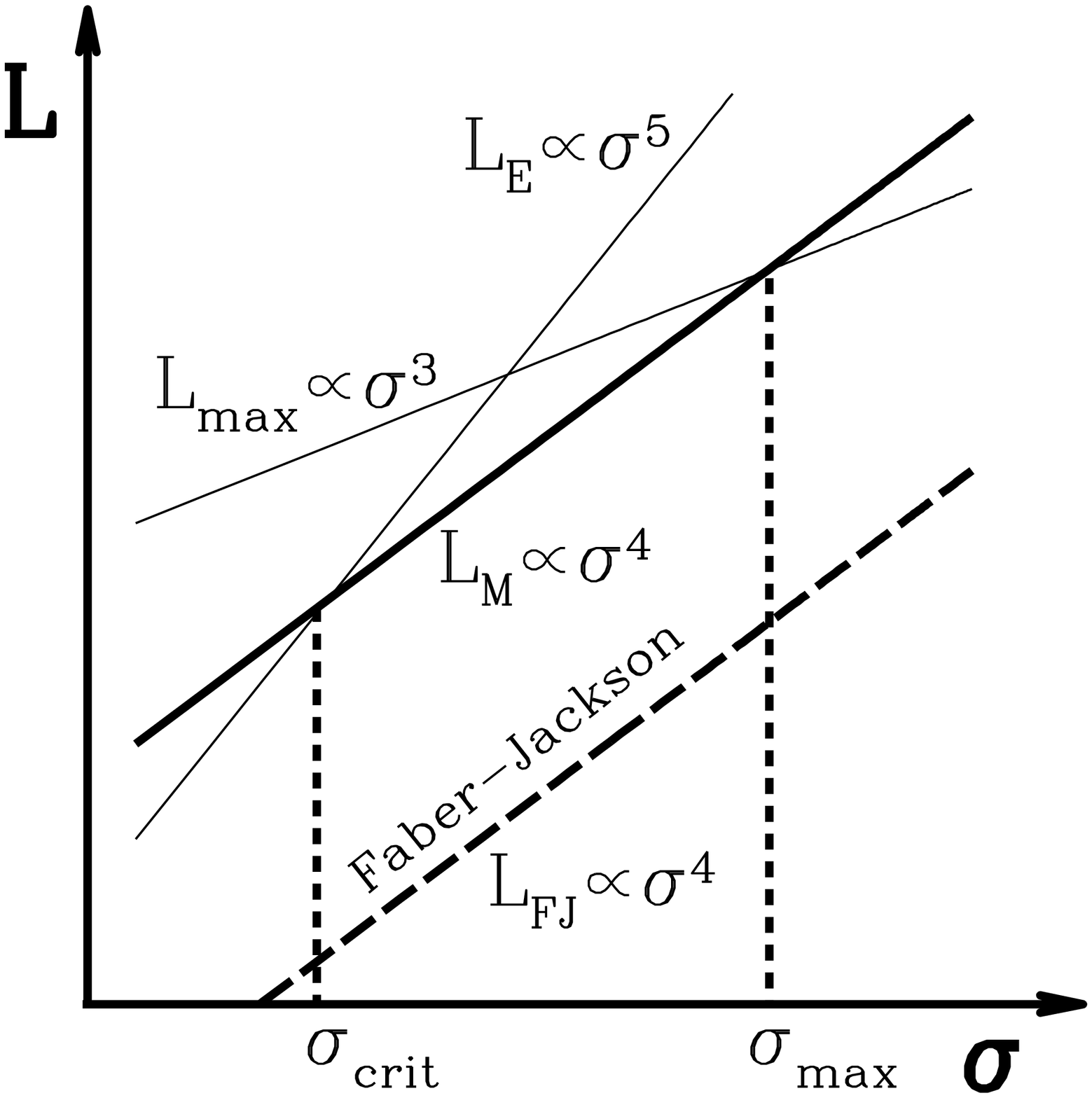}

\plotone{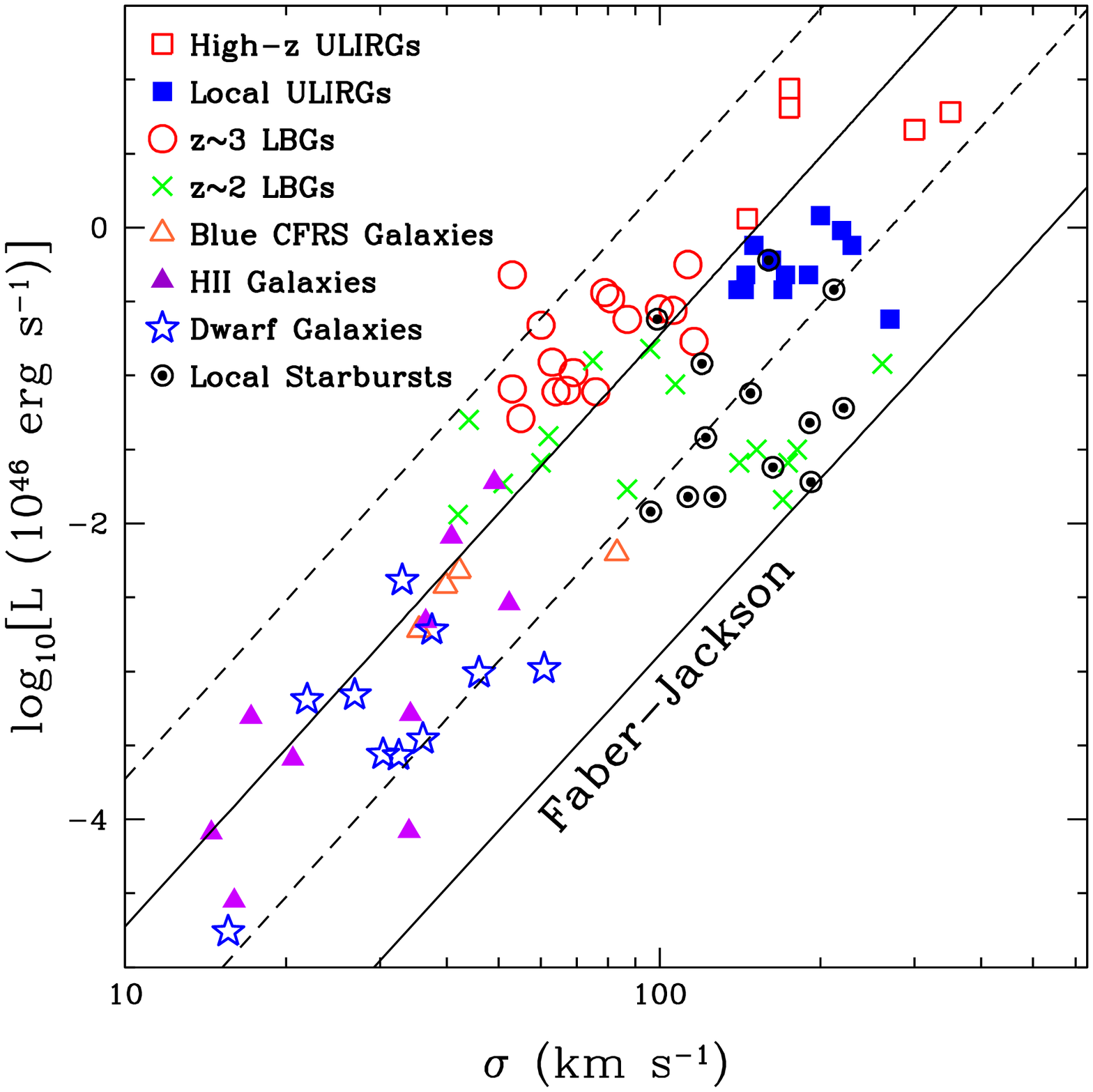}

\plotone{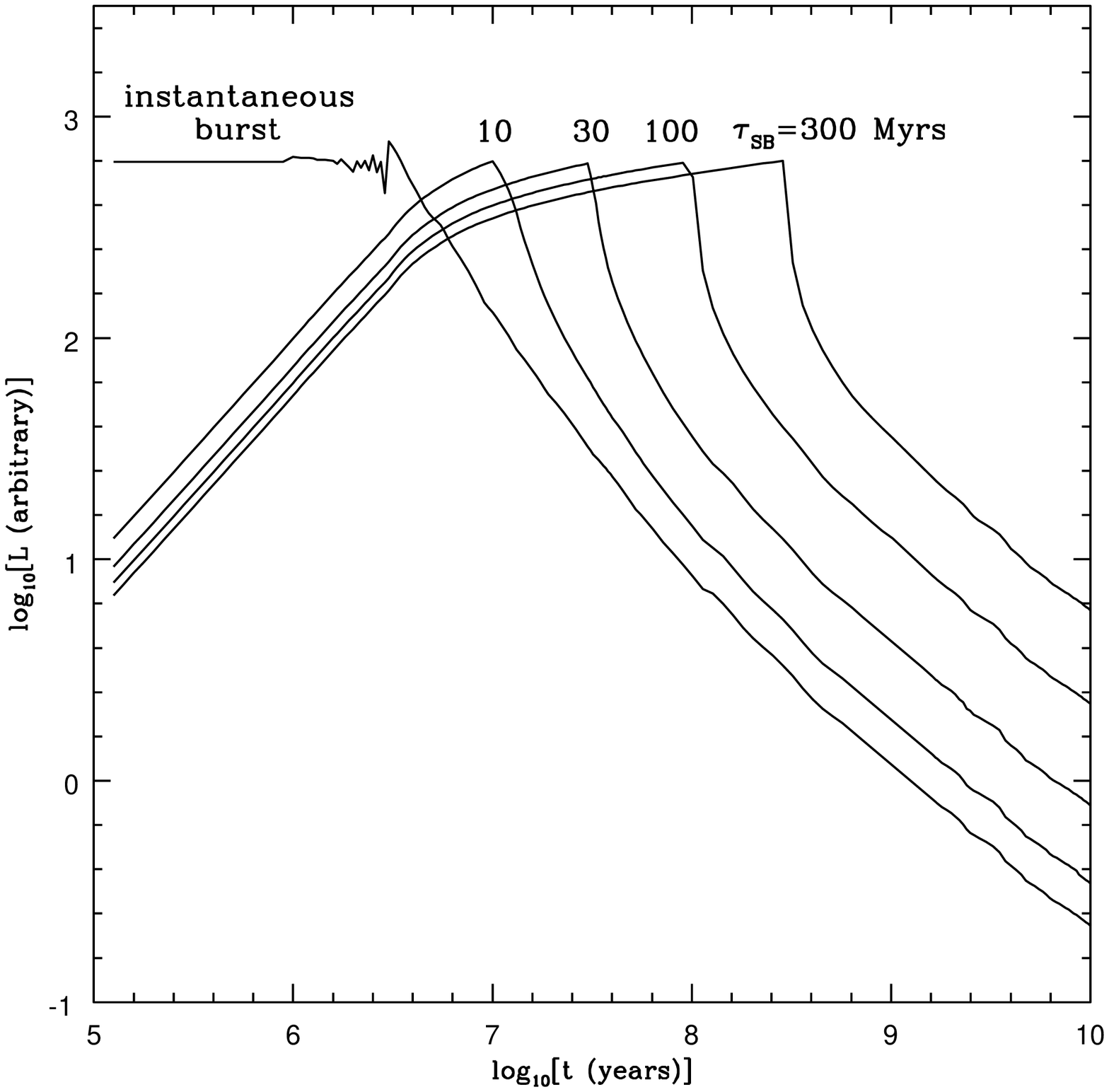}

\plotone{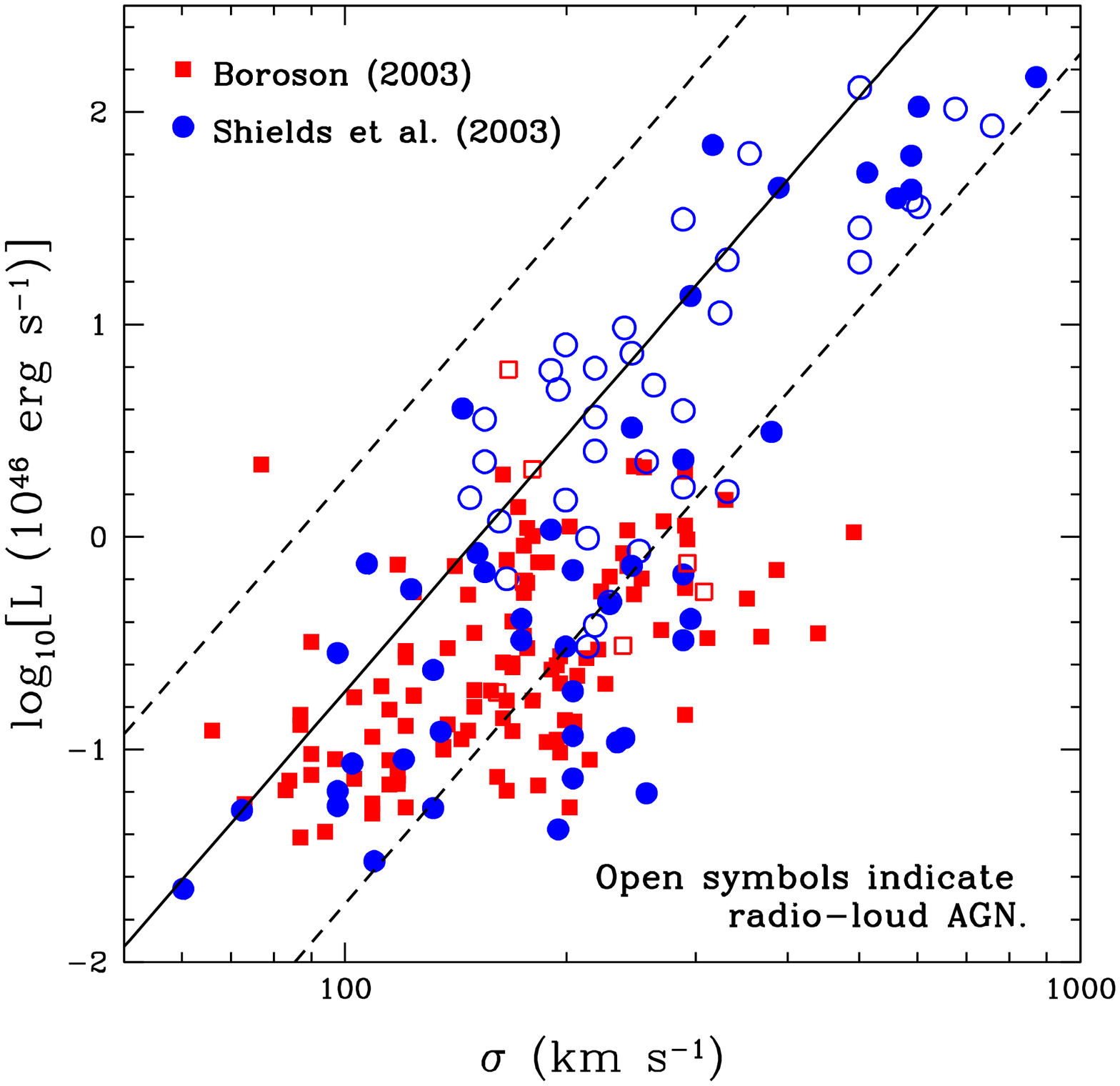}

\clearpage
\clearpage

\begin{table}
\begin{scriptsize}
\begin{center}
\caption{Luminosity \& Velocity Dispersion: High-Redshift Galaxies \label{tab:ls}}

\begin{tabular}{lcccccccc}
\hline \hline
\multicolumn{1}{c}{} &\multicolumn{1}{c}{} &\multicolumn{1}{c}{} &\multicolumn{1}{c}{} & \multicolumn{1}{c}{}\\

\multicolumn{1}{c}{Object Class} &\multicolumn{1}{c}{Name} & \multicolumn{1}{c}{$\log_{10}[L\,\,(L_\odot)]$}  &\multicolumn{1}{c}{$\sigma$ (km s$^{-1}$)} & \multicolumn{1}{c}{References \& Notes}\\

\multicolumn{1}{c}{} &\multicolumn{1}{c}{} & \multicolumn{1}{c}{ } &
\multicolumn{1}{c}{} &\multicolumn{1}{c}{} \\

\hline\hline

& &   \\

High-$z$ ULIRGs & SMM J14011+0252 & 13.36 & 175 & Tecza et al.~(2004); $L=L_{\rm IR}$.  $\sigma=V_{\rm Circ}/\sqrt{2}$.\\
&&&&&\\
		& SMM J02399-0136 & 13.08 & 300 & Genzel et al.~(2003); Neri et al.~(2003) \\
                & SMM J04431+0210 & 12.48 & 146 & For all systems:    \\
		& SMM J09431+4700 & 13.23 & 175 & $L=L_{\rm IR}$ and $\sigma={\rm LW}/2.4$, where \\
		& SMM J16368+4057 & 13.20 & 350 & ${\rm LW}=$ line width. \\

& &   \\
\hline
& &   \\

$z\sim3$ LBGs   & CDFa D18         & 11.98	& 79 &  Pettini et al.~(2001)  \\
                & CDFa C8          & 11.86	& 106 &  \\
                & CDFa C1	   & 11.51	& $\leq63$ &  	   For all systems:\\
                & Q0201+113 C6	   & 11.31	& 64 &  	   \\
                & Q0256-000 C17	   & 11.33	& 53 &  	   $L$ obtained from AB magnitude: \\
                & Q0347-383 C5	   & 11.44	& 69 &  	   $L=\nu L_{\nu_0}10^{-0.4M_{\rm UV}}$; \\
                & B2 0902+343 C12  & 11.80	& 87 &  	   $L_{\nu_0}=4.3\times10^{20}$ erg/s/Hz.\\
                & West MMD11	   & 12.10	& 53 &  	   \\
                & Q1422+231 D81	   & 11.65      & 116 &  	   Extinction $E(B-V)$ computed \\
		& 3C324 C3         & 11.31	& 76  &            from $(G-R)$ color\\
                & SSA22a MD46	   & 11.32      & 67 &  	   (A.~Shapley, private communication).\\
                & SSA22a D3	   & 12.17	& 113 &  	   \\
                & DSF 2237+116a C2 & 11.87	& 100 &  	   $\sigma=\,$1D velocity dispersion\\
                & B2 0902+343 C6   & 12.13	& 55 &  	   of nebular emission lines.\\
                & MS 1512-cB58	   & 11.94	& 81 &  	   \\
		& Q0000-263 D6	   & 11.76	& 60 &		   \\

& &   \\
\hline
& &   \\

$z\sim2$ LBGs   & CDFb-BN88	   & 11.60	& 96 &  Erb et al.~(2003) & \\
                & Q0201-B13	   & 11.01	& 62 &  	 \\
                & Westphal BX600   & 10.92	& 181 &  	  For all systems:\\
                & Q1623-BX376	   & 11.50	& 261(a), $<224$(b) &  	  \\
                & Q1623-BX432	   & 10.69	& 51 &  	 $L$ obtained from extinction corrected UV \\
                & Q1623-BX447	   & 10.83	& 174 &  	  SFR via $SFR=1.4\times10^{-28}L_{1500}$ erg/s/Hz. \\
                & Q1623-BX449	   & 10.83	& 141 &  	   $\sigma$ from their Table 2.\\
                & Q1623-BX511	   & 10.92	& 152 &  	  \\
                & Q1623-BX522	   & 11.12	& $<44$ &  	  For system Q1623-BX376,\\
                & Q1623-MD107	   & 10.48	& $<42$ &  	  we plot $\sigma=261$ km s$^{-1}$.\\
                & Q1700-BX691	   & 10.58	& 170 &  	  \\
                & Q1700-BX717	   & 10.83	& $<60$ &  	  The other systems with only upper\\
                & Q1700-MD103	   & 11.52	& 75 &  	  bounds on $\sigma$ we include in \\
                & Q1700-MD109	   & 10.65	& 87 &  	  Figure \ref{plot:ls}.\\
                & SSA22a-MD41	   & 11.36	& 107 &  	  \\


& &   \\
\hline
& &   \\

$z\sim0.6$ CFRS  & CFRS-1 & 9.70	& 35  &  Mall\'{e}n-Ornelas et al.~(1999)  \\
                 & CFRS-2 & 10.00	& 40 &    For all systems: \\
                 & CFRS-3 & 10.10	& 42  &    $L$ obtained from $M_B$ with $h=0.7$;\\
                 & CFRS-4 & 10.22	& 83  &     $L=4.73\,L_{B,\odot}\,10^{-0.4(M_B-5.48)}$.\\

& &   \\

\hline
\hline
\end{tabular}
\end{center}
\end{scriptsize}
\end{table}

\clearpage
\clearpage

\begin{table}
\begin{scriptsize}
\begin{center}
\caption{Luminosity \& Velocity Dispersion: Low-Redshift Galaxies \label{tab:ls3}}

\begin{tabular}{lcccccccc}
\hline \hline
\multicolumn{1}{c}{} &\multicolumn{1}{c}{} &\multicolumn{1}{c}{} &\multicolumn{1}{c}{} & \multicolumn{1}{c}{}\\

\multicolumn{1}{c}{Object Class} &\multicolumn{1}{c}{Name} & \multicolumn{1}{c}{$\log_{10}[L\,\,(L_\odot)]$}  &\multicolumn{1}{c}{$\sigma$ (km s$^{-1}$)} & \multicolumn{1}{c}{References \& Notes}\\

\multicolumn{1}{c}{} &\multicolumn{1}{c}{} & \multicolumn{1}{c}{ } &
\multicolumn{1}{c}{} &\multicolumn{1}{c}{} \\

\hline\hline

& &   \\

Local ULIRGs    & I00262+4251             & 12.0	& 170 &   Genzel et al.~(2001)  \\
                & I00456-2901		  & 12.2	& 162 &     \\
                & I01388-4618		  & 12.0	& 144 &     For all systems:\\
                & I01572+0009 Mrk 1014	  & 12.5	& 200 &     \\
                & I14348-1447		  & 12.3	& 150(u), 170(l) &    $L=L_{\rm IR}$; $\sigma=\sigma$ stars\\
                & I15327+2340 Arp 220     & 12.1	& 172(W), 155(E) &     (their Tables 1 and 2).\\
                & I16504+0228 NGC 6240    & 11.8	& 270(N), 290(S) &     \\
                & I17208-0014		  & 12.3	& 229  &    In cases where two $\sigma$s are \\
                & I20087-0308		  & 12.4	& 219  &    observed, we plot the lower \\
                & I20551-4250		  & 12.0	& 140 &     of the two.\\
                & I23365+3604		  & 12.1	& 145 &     \\
                & I23578-5307		  & 12.1	& 190  &    \\


& &   \\
\hline
& &   \\

HII Galaxies     & UM133          &  9.11	& 17.2	 &  Telles \& Terlevich (1997) \\
                 & C0840+1044     &  8.34	& 34.0   &   Melnick et al.~(1988)\\
                 & C0840+1201     &  9.76	& 36.5	 &    For all systems:  	\\
                 & C08-28A	  &  10.70	& 49.1	 &    \\
                 & Mrk 36	  &  7.87	& 16.0	 &   $L$ obtained from quoted $M_V$; \\
                 & UM 448	  &  10.33	& 40.8	 &    $L=6.3\,L_{V,\odot}\,10^{-0.4(M_V-4.83)}$.\\
                 & UM 455	  &  8.83	& 20.6	 &    \\
                 & UM 461A	  &  8.33	& 14.5	 &    Corrected for galactic extinction only.		\\
                 & C1212+1148     &  9.13	& 34.2	 &    	\\
                 & C1409+1200     &  9.88	& 52.3	 &    $\sigma$ quoted as rms velocity dispersion.	\\

& &   \\
\hline
& &   \\

Dwarf Galaxies       & WLM        & 7.66  &  15.6  &  Mateo (1998)  \\
                     & NGC 55	  & 9.44  &  60.8  &    For all systems: $L$ obtained from SFR  \\
                     & NGC 6822	  & 8.96  &  36.1  &    using $L=\epsilon\dot{M}_\star c^2$ with $\epsilon=10^{-3}$.\\
                     & IC 10	  & 10.03 &  33.0  &    $\sigma=V_{\rm Circ}/[\sqrt{2}\sin i]$\\
&&&&&\\
		     & II Zw 40   & 9.18  & 35.2	 &  Martin (1998)  \\
		     & NGC 1569	  & 9.23  & 21.9  &    For all systems: $L$ obtained from SFR \\
		     & NGC 4861	  & 8.86  & 30.4  &    using $L=\epsilon\dot{M}_\star c^2$ with $\epsilon=10^{-3}$.	\\
		     & NGC 1800	  & 8.85  & 32.5  &    $\sigma=V_{\rm Circ}/[\sqrt{2}]$	\\
		     & NGC 3077	  & 9.41  & 45.96  &    \\
		     & NGC 5253	  & 9.26  & 26.9$-$42.4	  &    For NGC 5253, we plot $\sigma=26.9$ km s$^{-1}$.\\

& &   \\

& &   \\
\hline
& &   \\

Local Starbursts & IRAS 03514+1546	& 11.1	& 191       & Heckman et al.~(2000)\\
 & NGC 1572	& 11.2	& 221       & \\
 & NGC 1614	& 11.3	& 148       & For all systems:\\
 & IRAS 04370-2416	& 11.0	& 122       & \\
 & NGC 1808	& 10.6	& 113       &  $L=L_{\rm IR}$; $\sigma=V_{\rm Rot}/\sqrt{2}$. \\ 
 & NGC 2146	& 10.7	& 192       & \\ 
 & M82	        & 10.5	& 96.2 &  $\sigma$ for M 82 from Martin (1998).\\
 & IRAS 10173+0828	& 11.8	& 99 & \\
 & NGC 3256	& 11.5	& 120 & \\
 & IRAS 10565+2448	& 12.0	& 212 & \\
 & Mrk 273	        & 12.2	& 160 & $\sigma$ for Mrk 273 from Genzel et al.~(2001).\\
 & NGC 7582	& 10.6	& 127 & \\
 & NGC 7552	& 10.8	& 163  & \\
& &   \\

\hline
\hline
\\
\end{tabular}
\end{center}
\end{scriptsize}
\end{table}


\begin{thebibliography}{}

\bibitem[Adelberger \& Steidel(2000)]{2000ApJ...544..218A}
  Adelberger, K.~L.~\& Steidel, C.~C.\ 2000, \apj, 544, 218

\bibitem[Adelberger et al.~(2003)]{adelberger}Adelberger, K.~L., Steidel, C.~C., Shapley, A.~E., \& Pettini, M.~2003, ApJ, 584, 45

\bibitem[Aguirre (1999)]{aguirre1999}Aguirre, A.~1999, ApJ, 525, 583

\bibitem[Aguirre et al.(2001)]{2001ApJ...561..521A} Aguirre, A.,
  Hernquist, L., Schaye, J., Katz, N., Weinberg, D.~H., \& Gardner,
  J.\ 2001a, \apj, 561, 521

\bibitem[Aguirre et al.(2001)]{2001ApJ...560..599A} Aguirre, A.,
  Hernquist, L., Schaye, J., Weinberg, D.~H., Katz, N., \& Gardner,
  J.\ 2001b, \apj, 560, 599 

\bibitem[Aguirre et al.(2001)]{2001ApJ...556L..11A} Aguirre, A.,
  Hernquist, L., Katz, N., Gardner, J., \& Weinberg, D.\ 2001c,
  \apjl, 556, L11 

\bibitem[Bernardi et al.(2003)]{2003AJ....125.1849B} 
 Bernardi, M., et al.\ 2003, \aj, 125, 1849 
 
\bibitem[Blandford(1999)]{1999gady.conf...87B} Blandford, R.~D.\ 1999, ASP  Conf.~Ser.~182: Galaxy Dynamics - A Rutgers Symposium, 87 

\bibitem[Boroson(2003)]{2003ApJ...585..647B} Boroson, T.~A.\ 2003, \apj, 
585, 647 

\bibitem[Bruzual A.~\& Charlot(2003)]{1993ApJ...405..538B} 
 Bruzual A., G.~\& Charlot, S.\ 2003, MNRAS, 344, 1000

\bibitem[Bullock et al.~(2001)]{bullock}
 Bullock, J.~S., Dekel, A., Kolatt, T.~S., Kravtsov, A.~V., Klypin, A.~A., 
 Porciani, C., \& Primack, J.~R.~2001, ApJ, 555, 240 
 
\bibitem[Burstein, Bender, Faber, \& Nolthenius(1997)]{1997AJ....114.1365B}  Burstein, D., Bender, R., Faber, S., \& Nolthenius, R.\ 1997, \aj, 114, 1365 

\bibitem[Calzetti(2001)]{2001PASP..113.1449C} Calzetti, D.\ 2001,
  \pasp, 113, 1449 

\bibitem[Chevalier \& Clegg(1985)]{1985Natur.317...44C} Chevalier,
  R.~A.~\& Clegg, A.~W.\ 1985, \nat, 317, 44 

\bibitem[Davies, Alton, Bianchi, \& Trewhella(1998)]{1998MNRAS.300.1006D} 
Davies, J.~I., Alton, P., Bianchi, S., \& Trewhella, M.\ 1998, \mnras, 300, 
1006 

\bibitem[Dekel \& Silk(1986)]{1986ApJ...303...39D} Dekel, A.~\& Silk,
  J.\ 1986, \apj, 303, 39 

\bibitem[De Young \& Heckman(1994)]{1994ApJ...431..598D} De Young,
  D.~S.~\& Heckman, T.~M.\ 1994, \apj, 431, 598 

\bibitem[Draine \& Lee(1984)]{1984ApJ...285...89D} Draine, B.~T.~\&
  Lee, H.~M.\ 1984, \apj, 285, 89 

\bibitem[Draine \& Salpeter(1979)]{1979ApJ...231...77D} Draine, B.~T.~\& 
Salpeter, E.~E.\ 1979, \apj, 231, 77

\bibitem[Elmegreen(1983)]{1983MNRAS.203.1011E} Elmegreen, B.~G.\ 1983, 
\mnras, 203, 1011 

\bibitem[Erb et al.(2003)]{2003ApJ...591..101E} 
Erb, D.~K., Shapley, A.~E.,  Steidel, C.~C., Pettini, M., Adelberger, K.~L., Hunt, M.~P., Moorwood,  A.~F.~M., \& Cuby, J.\ 2003, \apj, 591, 101  
  
\bibitem[Faber \& Jackson(1976)]{1976ApJ...204..668F} Faber, S.~M.~\&
 Jackson, R.~E.\ 1976, \apj, 204, 668

\bibitem[Fabian(1999)]{1999MNRAS.308L..39F} Fabian, A.~C.\ 1999,
  \mnras, 308, L39
  
\bibitem[Fabian et al.~(2002)]{fabian2002}
Fabian, A.~C., Wilman, R.~J., \& Crawford, C.~S.~2002, MNRAS, 329, L18

\bibitem[Ferrara \& Tolstoy(2000)]{2000MNRAS.313..291F} Ferrara, A.~\&
  Tolstoy, E.\ 2000, \mnras, 313, 291 

\bibitem[Ferrarese \& Merritt(2000)]{2000ApJ...539L...9F} Ferrarese,
  L.~\& Merritt, D.\ 2000, \apjl, 539, L9 

\bibitem[Gebhardt et al.(2000)]{2000ApJ...539L..13G} Gebhardt, K., et al.\ 2000, \apjl, 539, L13 
 
\bibitem[Genzel \& Cesarsky(2000)]{2000ARA&A..38..761G} Genzel, R.~\&  Cesarsky, C.~J.\ 2000, \araa, 38, 761   

\bibitem[Genzel et al.(2001)]{2001ApJ...563..527G} Genzel, R., Tacconi,  L.~J., Rigopoulou, D., Lutz, D., \& Tecza, M.\ 2001, \apj, 563, 527   

\bibitem[Genzel et al.(2003)]{2003ApJ...584..633G} Genzel, R., Baker,  A.~J., Tacconi, L.~J., Lutz, D., Cox, P., Guilloteau, S., \& Omont, A.\  2003, \apj, 584, 633  

\bibitem[Genzel et al.~(2004)]{genzel}
Genzel, R., et al.~2004,  to appear in the proceedings of the Venice 
conference "Multiwavelength Mapping of Galaxy Formation and Evolution". Astro-ph/0403183
 
\bibitem[Giovanelli et al.(1997)]{1997ApJ...477L...1G} Giovanelli, R.,  Haynes, M.~P., da Costa, L.~N., Freudling, W., Salzer, J.~J., \& Wegner,  G.\ 1997, \apjl, 477, L1   

\bibitem[Haehnelt(1995)]{1995MNRAS.273..249H} Haehnelt, M.~G.\ 1995,
  \mnras, 273, 249

\bibitem[Haehnelt, Natarajan, \& Rees(1998)]{1998MNRAS.300..817H}
  Haehnelt, M.~G., Natarajan, P., \& Rees, M.~J.\ 1998, \mnras, 300,
  817 

\bibitem[Hartwell et al.(2004)]{2004MNRAS.348..406H} Hartwell, J.~M.,
  Stevens, I.~R., Strickland, D.~K., Heckman, T.~M., \& Summers,
  L.~K.\ 2004, \mnras, 348, 406 

\bibitem[Heavens, Panter, Jimenez, \& Dunlop(2004)]{2004Natur.428..625H} Heavens, A., Panter, B., Jimenez, R., \& Dunlop, J.\ 2004, \nat, 428, 625 

\bibitem[Heckman, Armus, \& Miley(1990)]{1990ApJS...74..833H} Heckman,
  T.~M., Armus, L., \& Miley, G.~K.\ 1990, \apjs, 74, 833  

\bibitem[Heckman(2000)]{2000RSPTA.358.2077H} Heckman, T.~M.\ 2000, Royal  Society of London Philosophical Transactions Series A, 358, 2077 

\bibitem[Heckman, Lehnert, Strickland, \& Armus(2000)]{2000ApJS..129..493H}  Heckman, T.~M., Lehnert, M.~D., Strickland, D.~K., \& Armus, L.\ 2000,  \apjs, 129, 493  

\bibitem[Kaspi et al.(2000)]{2000ApJ...533..631K} Kaspi, S., Smith, P.~S., Netzer, H., Maoz, D., Jannuzi, B.~T., \& Giveon, U.\ 2000, \apj, 533, 631 

\bibitem[Kennicutt, Tamblyn, \& Congdon(1994)]{1994ApJ...435...22K}  Kennicutt, R.~C., Tamblyn, P., \& Congdon, C.~E.\ 1994, \apj, 435, 22 

\bibitem[King(2003)]{2003ApJ...596L..27K} King, A.\ 2003, \apjl, 596, L27 

\bibitem[Klein, McKee, \& Colella(1994)]{1994ApJ...420..213K} Klein, R.~I., McKee, C.~F., \& Colella, P.\ 1994, \apj, 420, 213 

\bibitem[Kozasa, Hasegawa, \& Nomoto(1989)]{1989ApJ...344..325K}
  Kozasa, T., Hasegawa, H., \& Nomoto, K.\ 1989, \apj, 344, 325 

\bibitem[Lamers \& Cassinelli (1999)]{lamers}
Lamers, H.~\& Cassinelli, J.~P., {\it Introduction to Stellar Winds}
(Cambridge University Press, Cambridge, 1999)

\bibitem[Lehnert \& Heckman(1996)]{1996ApJ...472..546L} Lehnert, 
 M.~D.~\& Heckman, T.~M.\ 1996, \apj, 472, 546 

\bibitem[Leitherer et al.(1999)]{1999ApJS..123....3L} Leitherer, C.,
  et al.\ 1999, \apjs, 123, 3 

\bibitem[Lilly, Le Fevre, Hammer, \& Crampton(1996)]{1996ApJ...460L...1L}  Lilly, S.~J., Le Fevre, O., Hammer, F., \& Crampton, D.\ 1996, \apjl, 460,  L1  

\bibitem[Mac Low \& Ferrara(1999)]{1999ApJ...513..142M} Mac Low, M.~\&
  Ferrara, A.\ 1999, \apj, 513, 142 

\bibitem[Madau et al.(1996)]{1996MNRAS.283.1388M} Madau, P., Ferguson,  H.~C., Dickinson, M.~E., Giavalisco, M., Steidel, C.~C., \& Fruchter, A.\  1996, \mnras, 283, 1388   

\bibitem[Madau, Pozzetti, \& Dickinson(1998)]{1998ApJ...498..106M} Madau, 
P., Pozzetti, L., \& Dickinson, M.\ 1998, \apj, 498, 106 

\bibitem[Mall{\' e}n-Ornelas, Lilly, Crampton, \&  Schade(1999)]{1999ApJ...518L..83M} Mall{\' e}n-Ornelas, G., Lilly, S.~J.,  Crampton, D., \& Schade, D.\ 1999, \apjl, 518, L83   

\bibitem[Martin(1998)]{1998ApJ...506..222M} Martin, C.~L.\ 1998, \apj, 506, 222 

\bibitem[Martin(1999)]{1999ApJ...513..156M} Martin, C.~L.\ 1999, \apj, 513,  156 

\bibitem[Martin(2004)]{martin2004} Martin, C.~L.\ 2004, submitted to ApJ

\bibitem[Mateo(1998)]{Mateo}Mateo, M 1998, \araa, 36, 435

\bibitem[Meurer et al.(1995)]{1995AJ....110.2665M} Meurer, G.~R.,
  Heckman, T.~M., Leitherer, C., Kinney, A., Robert, C., \& Garnett,
  D.~R.\ 1995, \aj, 110, 2665 

\bibitem[Meurer et al.(1997)]{meureretal} Meurer, G.~R.,
  Heckman, T.~M., Lehnert, M.~D., Leitherer, C., \& Lowenthal, J.~1997, \aj, 114, 54

\bibitem[Mihos \& Hernquist(1996)]{1996ApJ...464..641M} Mihos,
  J.~C.~\& Hernquist, L.\ 1996, \apj, 464, 641 

\bibitem[Mo, Mao, \& White(1998)]{1998MNRAS.295..319M} Mo, H.~J., Mao,  S., \& White, S.~D.~M.\ 1998, \mnras, 295, 319 

\bibitem[Nelson(2000)]{2000ApJ...544L..91N} Nelson, C.~H.\ 2000, \apjl, 
544, L91

\bibitem[Nelson \& Whittle(1996)]{1996ApJ...465...96N} Nelson, C.~H.~\& 
Whittle, M.\ 1996, \apj, 465, 96

\bibitem[Neri et al.(2003)]{2003ApJ...597L.113N} Neri, R., et al.\ 2003, \apjl, 597, L113 

\bibitem[Netzer \& Elitzur(1993)]{1993ApJ...410..701N} Netzer, N.~\&  Elitzur, M.\ 1993, \apj, 410, 701

\bibitem[Nozawa et al.(2003)]{2003ApJ...598..785N} Nozawa, T., Kozasa,
  T., Umeda, H., Maeda, K., \& Nomoto, K.\ 2003, \apj, 598, 785 

\bibitem[Nulsen et al.~(2004)]{nulsen}
Nulsen, P., McNamara, B., Wise, M., \& David, L.~2004, submitted to ApJ, Astro-ph/0408315

\bibitem[Pahre, Djorgovski, \& de Carvalho(1998)]{1998AJ....116.1591P} Pahre, M.~A., Djorgovski, S.~G., \& de Carvalho, R.~R.\ 1998, \aj, 116, 1591 

\bibitem[Pettini et al.(2000)]{2000ApJ...528...96P} Pettini, M., Steidel,  C.~C., Adelberger, K.~L., Dickinson, M., \& Giavalisco, M.\ 2000, \apj,  528, 96   

\bibitem[Pettini et al.(2001)]{2001ApJ...554..981P} 
Pettini, M., Shapley,  A.~E., Steidel, C.~C., Cuby, J., Dickinson, M., Moorwood, A.~F.~M.,  Adelberger, K.~L., \& Giavalisco, M.\ 2001, \apj, 554, 981   

\bibitem[Pierini \& Tuffs(1999)]{1999A&A...343..751P} Pierini, D.~\& Tuffs,  R.~J.\ 1999, \aap, 343, 751   

\bibitem[Poludnenko, Frank, \& Blackman(2002)]{2002ApJ...576..832P} 
Poludnenko, A.~Y., Frank, A., \& Blackman, E.~G.\ 2002, \apj, 576, 832

\bibitem[Ruszkowski \& Begelman(2002)]{2002ApJ...581..223R} Ruszkowski, 
M.~\& Begelman, M.~C.\ 2002, \apj, 581, 223 

\bibitem[Ruszkowski, Br{\" u}ggen, \& Begelman(2004)]{2004ApJ...611..158R} 
Ruszkowski, M., Br{\" u}ggen, M., \& Begelman, M.~C.\ 2004, \apj, 611, 158 

\bibitem[Sanders \& Mirabel(1996)]{1996ARA&A..34..749S} Sanders, D.~B.~\& Mirabel,
 I.~F.\ 1996, \araa, 34, 749 
 
\bibitem[Scannapieco \& Oh (2004)]{scannapieco_oh}
 Scannapieco, E.~\& Oh, P.~2004, submitted to ApJ, Astro-ph/0401087  
 
 \bibitem[Shapley et al.(2001)]{2001ApJ...562...95S} Shapley, A.~E.,  Steidel, C.~C., Adelberger, K.~L., Dickinson, M., Giavalisco, M., \&  Pettini, M.\ 2001, \apj, 562, 95   

\bibitem[Scoville et al.~(2001)]{scoville2001}
Scoville, N., Polletta, M., Ewald, S., Stolovy, S., Thompson, R., \& Rieke, M.~2001, AJ, 122, 3017

\bibitem[Scoville et al.~(2003)]{scoville2003}
Scoville, N.~2003, JKAS, 36,167

\bibitem[Shields et al.(2003)]{2003ApJ...583..124S} Shields, G.~A., 
Gebhardt, K., Salviander, S., Wills, B.~J., Xie, B., Brotherton, M.~S., 
Yuan, J., \& Dietrich, M.\ 2003, \apj, 583, 124 

\bibitem[Silk \& Rees(1998)]{1998A&A...331L...1S} 
Silk, J.~\& Rees, M.~J.\ 1998, \aap, 331, L1 
 
\bibitem[Smail, Ivison, \& Blain(1997)]{1997ApJ...490L...5S} 
Smail, I.,  Ivison, R.~J., \& Blain, A.~W.\ 1997, \apjl, 490, L5  
 
 
\bibitem[Steidel et al.(1996)]{1996ApJ...462L..17S} 
Steidel, C.~C.,  Giavalisco, M., Pettini, M., Dickinson, M., \& Adelberger, K.~L.\ 1996,  \apjl, 462, L17 
 
\bibitem[Steidel et al.(1999)]{1999ApJ...519....1S} 
Steidel, C.~C.,  Adelberger, K.~L., Giavalisco, M., Dickinson, M., \& Pettini, M.\ 1999,  \apj, 519, 1   

\bibitem[Strickland (2004)]{strickland2004}
  Strickland, D.~K.~2004, in the proceedings of IAU symposium 222: 
  {\it The Interplay among Black Holes, Stars and ISM in Galactic Nuclei}, held in Gramado, Brazil, March 1-5 2004. Eds. Th. Storchi Bergmann, L.C. Ho \& H.R. Schmitt, Astro-ph/0404316
   
\bibitem[Strickland \& Stevens(2000)]{2000MNRAS.314..511S} Strickland,  D.~K.~\& Stevens, I.~R.\ 2000, \mnras, 314, 511  
    
\bibitem[Tecza et al.(2004)]{2004ApJ...605L.109T} Tecza, M., et al.\ 2004,  \apjl, 605, L109    

\bibitem[Telles \& Terlevich(1997)]{1997MNRAS.286..183T} Telles, E.~\&  Terlevich, R.\ 1997, \mnras, 286, 183    

\bibitem[Thornton, Gaudlitz, Janka, \& Steinmetz(1998)]
 {1998ApJ...500...95T} Thornton, K., Gaudlitz, M.,
  Janka, H.-T., \& Steinmetz, M.\ 1998, \apj, 500, 95 

\bibitem[Todini \& Ferrara(2001)]{2001MNRAS.325..726T} Todini, P.~\&
  Ferrara, A.\ 2001, \mnras, 325, 726 

\bibitem[Tremaine et al.(2002)]{2002ApJ...574..740T} Tremaine, S., et al.\ 2002, \apj, 574, 740 

\bibitem[Van Dokkum et al.~(2004)]{vandokkum} Van Dokkum et al., 2004, ApJ in press (astro-ph/0404471)

\bibitem[Vestergaard(2004)]{2004ApJ...601..676V} Vestergaard, M.\ 2004,  \apj, 601, 676   

\end{thebibliography}
\end{document}